\documentclass[a4paper,11pt]{article}
\pdfoutput=1 

\usepackage{jcappub}
\usepackage{multirow}
\usepackage{subcaption}
\usepackage{float}

\DeclareRobustCommand{\rchi}{{\mathpalette\irchi\relax}}
\newcommand{\irchi}[2]{\raisebox{\depth}{$#1\chi$}}

\makeatletter
\def\@fpheader{\relax}
\makeatother

\title{A Quasi-Static Approach to Structure Formation in Black Hole Universes}

\author{Jessie Durk and}
\author{Timothy Clifton}

\affiliation{School of Physics \& Astronomy, Queen Mary University of London, UK}

\emailAdd{j.durk@qmul.ac.uk}
\emailAdd{t.clifton@qmul.ac.uk}

\abstract{Motivated by the existence of hierarchies of structure in the Universe, we present four new families of exact initial data for inhomogeneous cosmological models at their maximum of expansion. These data generalise existing black hole lattice models to situations that contain clusters of masses, and hence allow the consequences of cosmological structures to be considered in a well-defined and non-perturbative fashion. The degree of clustering is controlled by a parameter $\lambda$, in such a way that for $\lambda \sim 0$ or $1$ we have very tightly clustered masses, whilst for $\lambda \sim 0.5$ all masses are separated by cosmological distance scales. We study the consequences of structure formation on the total net mass in each of our clusters, as well as calculating the cosmological consequences of the interaction energies both within and between clusters. The locations of the shared horizons that appear around groups of black holes, when they are brought sufficiently close together, are also identified and studied. We find that clustering can have surprisingly large effects on the scale of the cosmology, with models that contain thousands of black holes sometimes being as little as 30\% of the size of comparable Friedmann models with the same total proper mass. This deficit is comparable to what might be expected to occur from neglecting gravitational interaction energies in Friedmann cosmology, and suggests that these quantities may have a significant influence on the properties of the large-scale cosmology.}

\begin{document}
    
\maketitle
\flushbottom

\section{Introduction}

The standard concordance model of $\Lambda$CDM cosmology is based on the assumption that the Universe can be described at all points in space, and on all scales, as being close to a single, universal Friedmann-Lema\^{i}tre-Robertson-Walker (FLRW) solution of Einstein's field equations. Structures are then incorporated within this model by performing perturbative expansions around this very simple background geometry. Despite the great successes that this approach has enjoyed, there are both observational and theoretical reasons to question its validity. These arise principally from the fact that below $\sim 100$Mpc the matter content of the Universe is very inhomogeneous, and that a single perturbed FLRW solution may not be sufficient to model all aspects of the complex gravitational physics that could result from this structure. 

On the theoretical side, the root of the potential problems lies in the difficulties associated with the coarse-graining of space-time in Einstein's theory \cite{rvdh}, and the lack of any positive proof that all space-times that are {\it statistically} homogeneous and isotropic on sufficiently large scales should behave in the same way as the {\it exactly} homogeneous and isotropic dust-filled FLRW solutions \cite{buchert,chris}. Observationally, there also appears to be growing tension between the astronomical data collected from our local region of space \cite{H0local}, and those collected over distance scales of the entire observable Universe \cite{H0CMB}. Both of these facts motivate the study of inhomogeneous cosmological solutions of Einstein's field equations, in order to determine the extent to which the FLRW models accurately represent the large-scale properties of more general space-times, and whether or not it is possible to identify any new phenomena that could have consequences for cosmology.

The difficulties involved in solving Einstein's equations make it extremely difficult to find realistic geometries, capable of modelling all aspects of the complex web of structure we see around us. One is therefore usually required to either resort to perturbative methods \cite{goldberg, adamek}, numerical solutions \cite{bb,mp,gs}, or to try to make as much progress as possible using exact methods \cite{exact}. One set of models that provides an interesting interplay of all of these approaches are the black hole lattices, studied perturbatively in Refs. \cite{bruneton,bruneton2,cwa,vij,vij2,light2,magweyl}, numerically in Refs. \cite{yoo,evolution2,yoo2,evolution1,bent1,bent3,bent4,bent5}, and using exact methods in Refs. \cite{tim,evolution3,silence,tim3,matterincosmo,jess1,bounce,jess2}. The basic idea behind these models is to split the matter content of the Universe into discretised packets, which are then modelled as black holes. The result is a set of space-times that explicitly satisfy the requirement of being highly inhomogeneous on small scales, while still being approximately homogeneous and isotropic on large scales. 

Our aim is to perform a controlled investigation of the effects that the clustering of masses has on the scale and properties of the cosmology, within this class of models. This work builds on the pioneering studies of Lindquist, Wheeler and Misner who constructed the initial vacuum data for $n$ black holes at a moment of time-reversal symmetry \cite{lindquist,misner}. It also extends more recent work that has studied the properties of cosmological models that contain regular arrays of black holes positioned on a three-sphere \cite{tim}. While such situations are far too simple to model the full hierarchy of structures in the real Universe, they do provide a well-defined way to study some of the questions involving coarse-graining and average expansion outlined above. The approach we will follow in this paper adds an extra level of structure to these models, so that they essentially have three length scales: (i) the curvature radius of each of the black holes, (ii) the radius of a cluster of black holes, and (iii) the cosmological curvature scale.

The growth of structure has long been speculated to be a potential cause of deviations from the predictions of FLRW cosmology in the real Universe \cite{rasanen}, and our models provide a type of quasi-static approximation that could be used to test some of these ideas within the context of exact solutions to Einstein's field equations at moments of instantaneous staticity. This study builds on the work involving hierarchical structures that was recently performed in a similar context in Ref. \cite{korz2}, as well as the random distributions of black holes that were considered in Ref. \cite{korz1}. As with these previous studies, we find situations in which hierarchies of structures can significantly alter the large-scale properties of the Universe. We interpret these findings in terms of the gravitational interaction energies between massive bodies, and the effects that they have on cosmology.

This paper is organised as follows: in Section \ref{sec:initial} we review the initial data problem for a universe containing $n$ black holes. In Section \ref{sec:structure} we describe the method used to split each mass up into a cluster of masses. Section \ref{sec:horizons} then contains calculations of the positions of the apparent horizons of individual black holes, as well as those belonging to clusters of black holes. The results of this are then used to calculate three different mass parameters in Section \ref{sec:mass}, and to derive expressions for the interaction energies between black holes in Section \ref{sec:int}. In Section \ref{sec:scales} we compare our black hole universe models to their FLRW counterparts, and to each other, in order to investigate the effect of the clustering of masses. Finally, we conclude in Section \ref{sec:conclude}. Geometrised units are used throughout the entirety of this paper, such that $c = G = 1$. Spatial indices are denoted by Latin indices $(i, j, \dots)$, and space-time coordinates are denoted by Greek indices $(\mu, \nu, \dots)$.

\section{Initial Data for a Black Hole Universe}
\label{sec:initial}

This section reviews and extends the initial data problem studied in Ref. \cite{tim}. In general, the Einstein field equations can be solved by performing a $3 + 1$ decomposition on the metric, and by separating the field equations into sets of constraint and evolution equations. In vacuum, the intrinsic geometry and extrinsic curvature of an initial hypersurface then provide sufficient information to perform a unique evolution. This means that if we can solve the constraint equations at some initial time, then we have enough information to determine the geometry of the entire space-time. In the standard notation, these constraint equations read
\begin{subequations}
\begin{align}\label{eq:constraint1}
      \mathcal{R} + K^2 - K_{ij}K^{ij} & = 0\\
    \label{eq:constraint2}   D_j(K_i^{\,\,j} - \delta_i^{\,\,j}K) & = 0 \, ,
\end{align}
\end{subequations}
where $\mathcal{R}$ is the Ricci scalar of the intrinsic geometry of the initial hypersurface, $K_{ij}$ is the extrinsic curvature of this surface, and $D_j$ is the covariant derivative with respect to the metric of this 3-space, $g_{ij}$. 

\subsection{Time-reversal symmetry}

The extrinsic curvature $K_{ij}$ can then be written as the time derivative of the metric $g_{ij}$ as follows:
\begin{equation}
\label{eq:ext}
 K_{ij} = -\frac{1}{2}\partial_t g_{ij} \, ,
\end{equation}
where $\partial /\partial t$ is a vector orthogonal to the initial hypersurface. If we choose to consider hypersurfaces that are symmetric under time-reversal, and therefore instantaneously static, then Eq. \eqref{eq:ext} implies that the extrinsic curvature of this surface must vanish. As a result, Eq. \eqref{eq:constraint2} is automatically satisfied and Eq. \eqref{eq:constraint1} reduces to $\mathcal{R} = 0$.  If we now conformally rescale our metric, so that $g_{ij}=\psi^4 \tilde{g}_{ij}$, then we find that this result can be written as
\begin{equation}
\mathcal{R} = \psi^{-4}\tilde{\mathcal{R}} - 8 \psi^{-5} \tilde{D}^2 \psi = 0 \, ,
\end{equation}
where $\tilde{\mathcal{R}}$ is the scalar curvature of the conformal space, and $\tilde{D}_i$ is a covariant derivative on the conformal 3-space. If we now choose $\tilde{g}_{ij}$ to be the round metric of a 3-sphere, then we find $\tilde{\mathcal{R}} = 6$. We therefore have that both Eq. \eqref{eq:constraint1} and \eqref{eq:constraint2} are satisfied at moments of time-reversal symmetry if
\begin{equation}
\label{eq:conformal}
ds^2 = \psi^4(d\rchi^2 + \sin^2\rchi d\Omega^2) \, ,
\end{equation}
where $d\Omega^2 = d\theta^2 + \sin^2\theta d\phi^2$, and if $\psi$ satisfies
\begin{equation}
\label{eq:linear}
\tilde{D}^2 \psi = \frac{3}{4} \psi  \, .
\end{equation}
This equation is linear in $\psi$, and has the known solution $\psi \propto 1/\sin(\rchi/2)$. We can therefore linearly add any number of similar terms to gain a new solution of the constraint equations. We will find that each such term in this sum corresponds to a new black hole, positioned at a new location on the conformal 3-sphere. The data described above corresponds to the maximum of expansion of a cosmological model.

\subsection{Multiple Schwarzschild-like black holes}

A spatial slice through the Schwarzschild solution for a mass $m$, and in isotropic coordinates, can be written as
\begin{equation}
  \label{eq:schsol}
ds^2 = \left(1+ \frac{m}{2r}\right)^4(dr^2 + r^2 d\Omega^2) \, .
\end{equation}
If transformed into hyperspherical polar coordinates, using $r=\frac{m}{2} \tan (\frac{\rchi}{2})$, then this becomes
\begin{equation}\label{eq:schsol2}
ds^2= \left(\frac{\sqrt{m}}{2\sin({\rchi}/2)} +\frac{\sqrt{m}}{2\cos({\rchi/2})}\right)^4 (d \chi^2 + \sin^2 \rchi d \Omega^2) \, ,
\end{equation}
which is manifestly of the same form as Eq. \eqref{eq:conformal}. The first of the two terms in the conformal factor of this geometry diverges at $\rchi =0$, and the second diverges at $\rchi= \frac{\pi}{2}$. These two terms clearly satisfy Eq. (\ref{eq:linear}), both individually and as a sum. They are, of course, also related to each through the transformation $\rchi \rightarrow \pi- \rchi$, which corresponds to a rotation of the 3-sphere by an angle $\pi$.

We can generate any number of additional terms in the conformal factor $\psi$ by rotating the 3-sphere by any arbitrary angle, and by adding a new term of the form $1/\sin(\rchi/2)$ in the new coordinates that result. Moreover, we can exploit the linearity of Eq. (\ref{eq:linear}) by summing any $n$ such terms together, in order to obtain a new solution. This generates the conformal factor $\psi$ as follows:
\begin{equation} \label{psi0}
\psi(\rchi, \theta, \phi) = \sum_{i=1}^{n} \frac{\sqrt{\tilde{m}_i}}{2f_i(\rchi, \theta, \phi)} \, ,
\end{equation}
where $n$ is the total number of terms, and $\tilde{m}_i$ are a set of arbitrary constants (labelled in analogy to the mass parameters in Eq. \eqref{eq:schsol2}). The $f_i$ in this equation are the source functions
\begin{equation}
f_i = \sin\left({\frac{1}{2}\arccos(h_i)}\right) \, ,
\end{equation}
and the $h_i$ are given by
\begin{equation}\label{eq:functions}
h_i = w_i \cos{\rchi} +  x_i \sin{\rchi}\cos{\theta} + y_i \sin{\rchi}\sin{\theta}\cos{\phi} + z_i  \sin{\rchi}\sin{\theta}\sin{\phi} \, ,
\end{equation}
for a term that diverges at the position $(w_i, x_i, y_i, z_i)$. Each such position corresponds to the location of a point-like mass in the initial data, and should be accounted for by including a corresponding term in $\psi$. Any number of masses can be included in the model in this way.

The proof of this is as follows: consider rotating the lattice such that one of the masses appears at position $(1,0,0,0)$. Then, the source function for this mass is $h_1 = \cos{\rchi}$ so that $f_1 = \sin{\frac{\rchi}{2}}$. Recall now the transformation between Cartesian coordinates in a 4-dimensional Euclidean embedding space and hyperspherical coordinates on a unit 3-dimensional sphere within that space:
\begin{subequations}\label{eq:transformations}
\begin{align}
    w &= \cos\rchi, \\
    x &= \sin\rchi\cos\theta,\\
    y &= \sin\rchi\sin\theta\cos\phi,\\
    z &= \sin\rchi\sin\theta\sin\phi.
\end{align}
\end{subequations}
This must then also mean that $w = h_1$. Any subsequent $i$th mass can be rotated to be at position $(1,0,0,0)$, and in the coordinates after the rotation must have $w'=h_i$. The general matrix transformation for this rotation is of course,
\begin{equation}
\begin{pmatrix}
w' \\
x'\\
y'\\
z'\\
\end{pmatrix} =
\begin{pmatrix}
\alpha_i & \beta_i & \gamma_i &  \delta_i\\
- & - & - & -\\
- & - & - & -\\
- & - & - & -\\
\end{pmatrix}
\begin{pmatrix}
w \\
x\\
y\\
z\\
\end{pmatrix}
\end{equation}
where for ease of presentation we have neglected to include any terms beyond the top line of the matrix. Multiplying out we have
\begin{equation} \label{line1}
w' = \alpha_i w +  \beta_i x + \gamma_i y + \delta_i z\, ,
\end{equation}
which can be compared to the requirement $w_i^2 + x_i^2 + y_i^2 + z_i^2 =1$ for the position of the $i$th mass on the unit 3-sphere. For both of these equations to be true when $\{w,x,y,z\}=\{w_i,x_i,y_i,z_i\}$, which means when $w'=1$, requires $\{ \alpha_i ,\beta_i , \gamma_i,\delta_i \} = \{w_i,x_i,y_i,z_i\}$. Using this result together with Eqs. \eqref{eq:transformations} and \eqref{line1}  then gives 
\begin{equation}
w' =w_i \cos{\rchi} +  x_i \sin{\rchi}\cos{\theta} + y_i \sin{\rchi}\sin{\theta}\cos{\phi} + z_i  \sin{\rchi}\sin{\theta}\sin{\phi} \, ,
\end{equation}
which upon recalling $w'=h_i$ can be seen to give Eq. \eqref{eq:functions} directly. This completes the proof.

\section{Clusters of Black Holes}
\label{sec:structure}

We can construct exact initial data for a cosmological model containing $n$ black holes by using the prescription in the previous section. If we wanted these black holes to be equidistant from each of their nearest neighbours, so that they formed a perfectly regular lattice, then we could proceed by tiling the conformal 3-space with regular polyhedra and placing a Schwarzschild mass at the centre of each \cite{tim}. Due to the fact that there are only a finite number of ways of tiling a 3-sphere with regular polyhedra, it turns out that there are a limited number of black holes that can be included in a perfectly regular universe of this type: either $5$, $8$, $16$, $24$, $120$ or $600$ masses \cite{coxeter}. In what follows, we will generalise this set of models by relaxing the assumption of perfectly equidistant spacing. This will allow for larger numbers of black holes to be considered, and for the effects of clustering of the black holes to be studied.

\subsection{Extended lattice solutions}
\label{sec:ext}

Our starting point for creating clusters of black holes will be the six perfect lattices described above. To create the clusters we will take each of the black holes in the regular lattice models, and explode them into a number of new black holes. We will, however, do this in a very specific way, in order to preserve the result that the large-scale properties of the space-time should still be statistically homogeneous and isotropic (after suitable coarse-graining). We will split each mass in the regular arrangement, which were originally at the centre of each polyhedral cell, into $C_n$ new masses, where $C_n$ is the number of vertices of one of the cells of the original lattice. We will then move these new black holes along trajectories that start at the cell centre, and extend outwards towards each of the vertices. This will allow us to control the scale of each of the clusters by deciding how far along each of these trajectories we wish to position each of our new black holes.

\begin{figure}[t!]
    \centering
    \begin{subfigure}[b]{0.32\textwidth}
        \includegraphics[width=\textwidth]{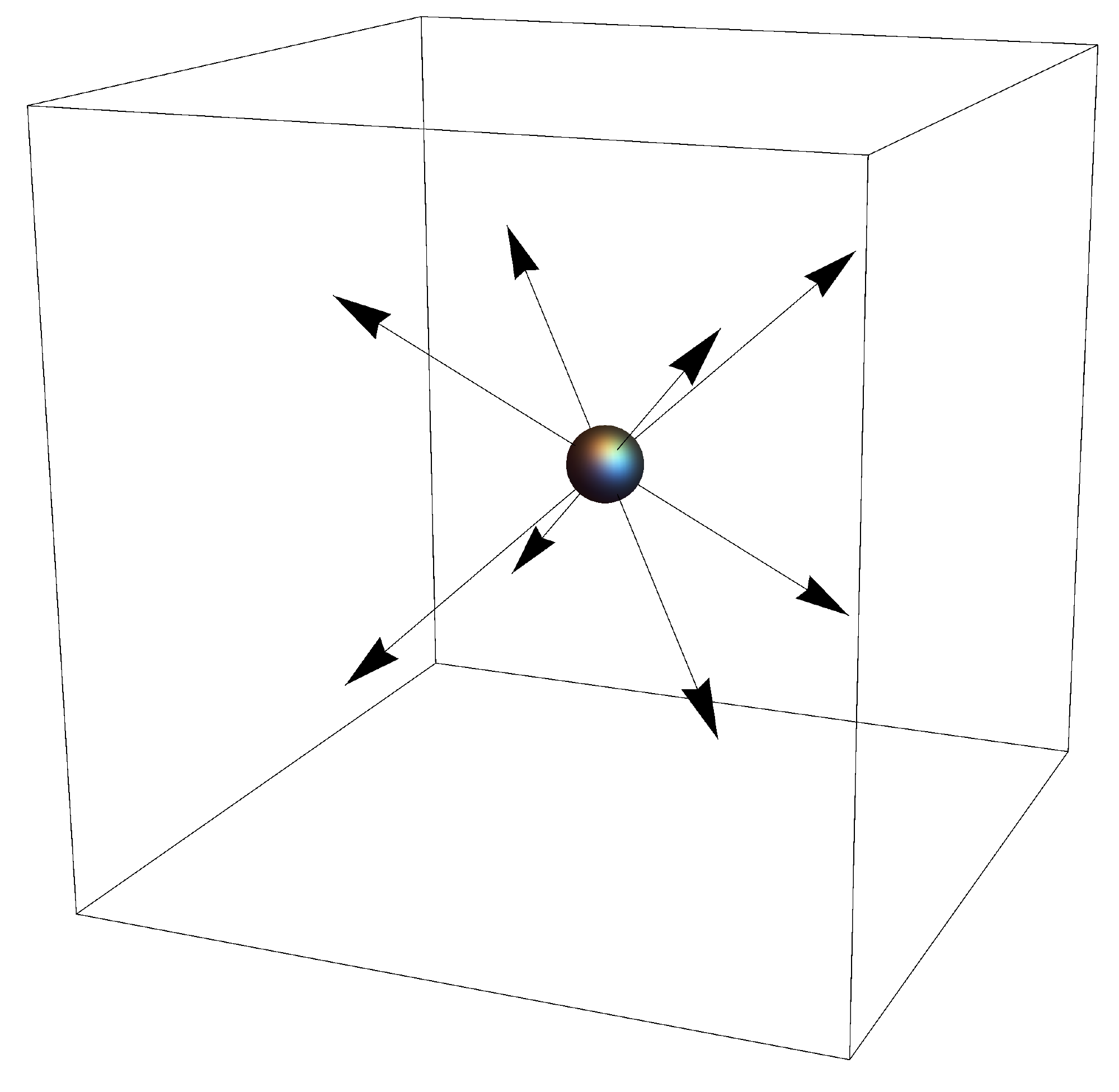}
        \caption{\centering{$\lambda=0$.}}
    \end{subfigure}
    \vspace{2mm}
    \begin{subfigure}[b]{0.32\textwidth}
        \includegraphics[width=\textwidth]{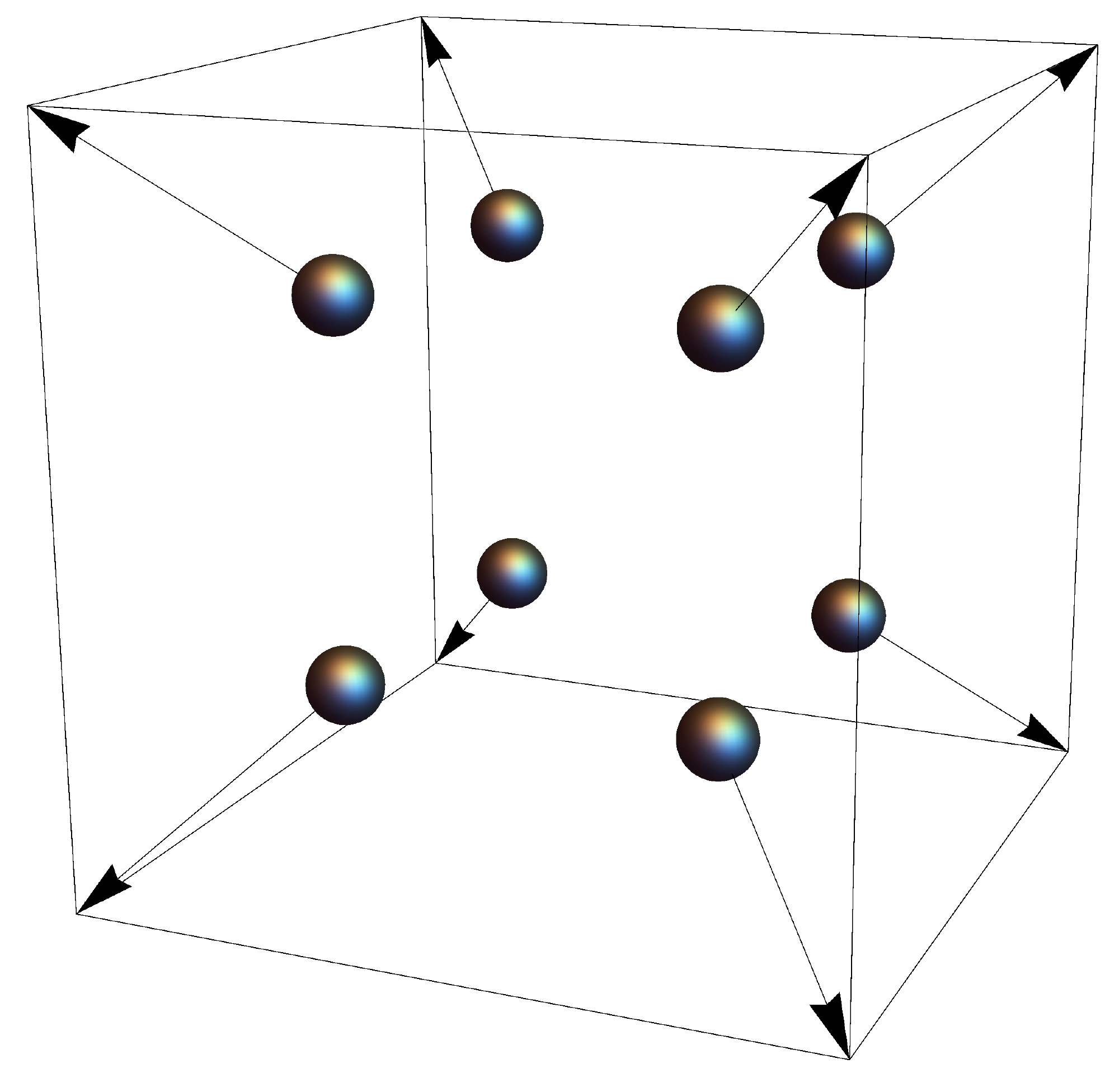}
           \caption{\centering{$\lambda=0.5$.}}
    \end{subfigure}
    \vspace{2mm}
    \begin{subfigure}[b]{0.32\textwidth}
        \includegraphics[width=\textwidth]{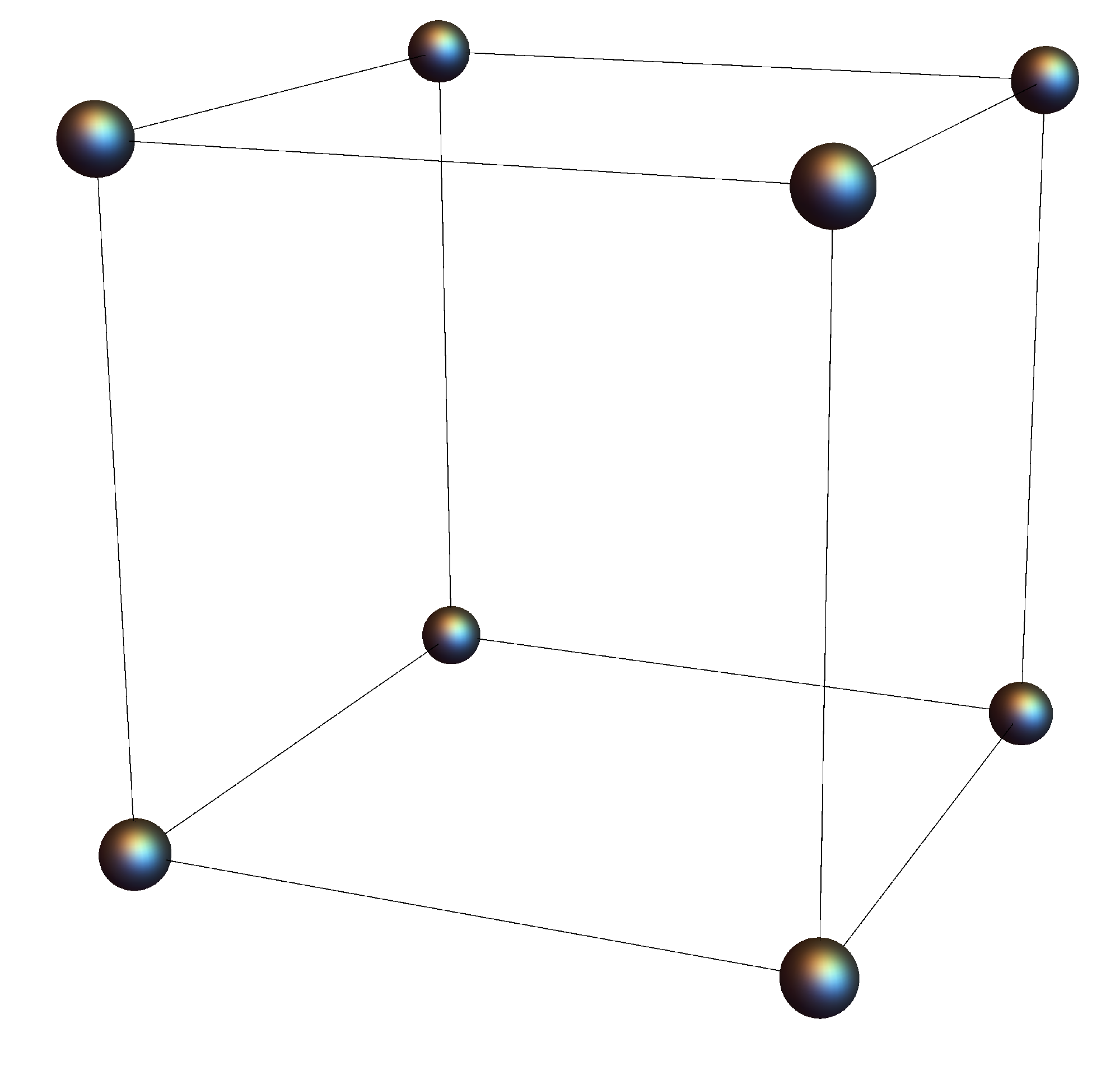}
           \caption{\centering{$\lambda=1$.}}
    \end{subfigure}
    \vspace{2mm}
 \caption{These images illustrate: (a) A single mass at the centre of a cubic cell in the 8-mass model, (b) the 8 new black holes that result from splitting up the original black hole that have then been moved part of the way towards the vertices, and (c) the final positions of the black holes after having been moved all the way to the vertices, and after having merged with black holes from the 4 other cells that meet at each vertex. The value of $\lambda$ here parameterises the distance the black holes have been moved, between the centre of a cell and the vertices.}
    \label{fig:cube}
\end{figure}

To illustrate this further, consider the regular 5-mass model that results from tiling the 3-sphere with 5 tetrahedra. Each of these tetrahedra has 4 vertices, which means each of the 5 masses will be split into 4 new masses, giving a total of 20 black holes in the new model. Each of the 4 black holes in each of the original primitive cells of the lattice will then be moved from the centre of the tetrahedron to one of the vertices. Upon reaching the end point of their trajectories, it will be the case that 4 masses will meet at each and every one of these vertices, as there are 4 cells meeting at every vertex point in this lattice. The model will then reduce back down to a 5 mass model, with each of the final black holes being located at what was previously a vertex point (which was originally the furthest point from any of the black holes). The fact that the 5-mass model contains 5 black holes both before and after this process is a reflection of the fact that this particular lattice is \textit{self-dual}. Starting from the 8-mass model this is not the case. For the 8-mass model each original black hole is split into 8 new black holes, as the primitive cell shape in this case is a cube. This results in a total of 64 new black holes, which after moving to the vertices of the cubic cells results in a lattice that eventually contains 16 black holes, as 4 cubic cells meet at each of the vertex points of a lattice of this type. The 8 and 16-mass models are thus said to be {\it duals} of each other. A schematic diagram of this particular process is shown in  Figure \ref{fig:cube} for a single cube, where a mass at the centre of the original lattice cell is split up into 8 masses, which move towards the vertices of the cube. For the remaining lattices, the 24-mass model can be seen to be self-dual, and contains a total of 144 new black holes after the splitting and before the merging. Finally, the 120 and 600-mass models are dual to each other, and contain a total of 2400 new black holes while the new black holes are in transit along our trajectories. Table \ref{tab:i} summarises these configurations for each of the original regular lattice models. This method generalises the work in Ref. \cite{tim}, as the regular lattices become the limiting case when the masses are either at the centre or the vertices of the original lattice cell.

\begin{table}[t!]
\centering
\begin{tabular}{|c|c|c|c|c|}
\hline
& & & &\\[-5pt]
Original $N^{\underline{o}}$ & Original & $N^{\underline{o}}$  of vertices & Total $N^{\underline{o}}$ of & $N^{\underline{o}}$ of black holes \\
of black holes & cell shape & of each cell & new black holes & in the dual lattice \\[5pt]
 \hline
 & & & &\\[-5pt]
5 & Tetrahedron & 4 & 20 & 5 \\
8 & Cube & 8 & 64 & 16\\
16 & Tetrahedron & 4 & 64 & 8\\
24 & Octahedron & 6 & 144 & 24\\
120 & Dodecahedron & 20 & 2400 & 600\\
600 & Tetrahedron & 4 & 2400 & 120\\[5pt]
\hline
\end{tabular}
\caption{\label{tab:i}The numbers of masses in each of the original regular lattices, the cell shapes of those lattices, the number of vertices of each cell, the total number of new black holes (after exploding the originals), and the number of black holes that result after they are moved all the way to the cell vertices.}
\end{table}

\noindent
Let us now consider how this process works in more detail, for each case:

\paragraph{5-Mass to 5-Mass Model:} The 5-mass model can be thought of as a regular tetrahedron along with an extra vertex added at a position that is equidistant from the existing four. The Cartesian coordinates, $(w,x,y,z)$, that give the positions of the masses in this model in the embedding space $E^4$ are
\begin{subequations}\label{eq:orig}
\begin{align}
i) & \quad (1,0,0,0) \\
ii)& \quad \sqrt{\frac{5}{16}}\left(-\frac{1}{\sqrt{5}},1,1,1\right)\\
iii)& \quad \sqrt{\frac{5}{16}}\left(-\frac{1}{\sqrt{5}},1,-1,-1\right)\\
iv)& \quad \sqrt{\frac{5}{16}}\left(-\frac{1}{\sqrt{5}},-1,1,-1\right)\\
v)& \quad \sqrt{\frac{5}{16}}\left(-\frac{1}{\sqrt{5}},-1,-1,1\right) \, .
\end{align}
\end{subequations}
The last four masses, labelled $ii)$ to $v)$, are at the locations of the vertices of a tetrahedron centred at $(1,0,0,0)$. Hence, with the addition of mass $i)$ at $(1,0,0,0)$, the 5-mass model is complete.

The positions of the masses of the dual lattice are straightforward to find in this case: one simply reflects the positions of the 5 locations above in the four planes given by $w=0$, $x=0$, $y=0$ and $z=0$. This can be achieved by simply putting a minus sign in front of each coordinate positions above, and the result gives the locations of the new black holes after they have been moved all the way along their trajectories. In Section \ref{sec:traj}, below, we give the formula we use for locating each of the black holes at positions between these initial and final states. The symbol $\lambda$ will be used to parameterise positions along these trajectories, such that $\lambda=0$ corresponds to the initial location and $\lambda =1$ corresponds to the final location.

\begin{figure}[t!]
    \centering
    \begin{subfigure}[b]{0.32\textwidth}
        \includegraphics[width=\textwidth]{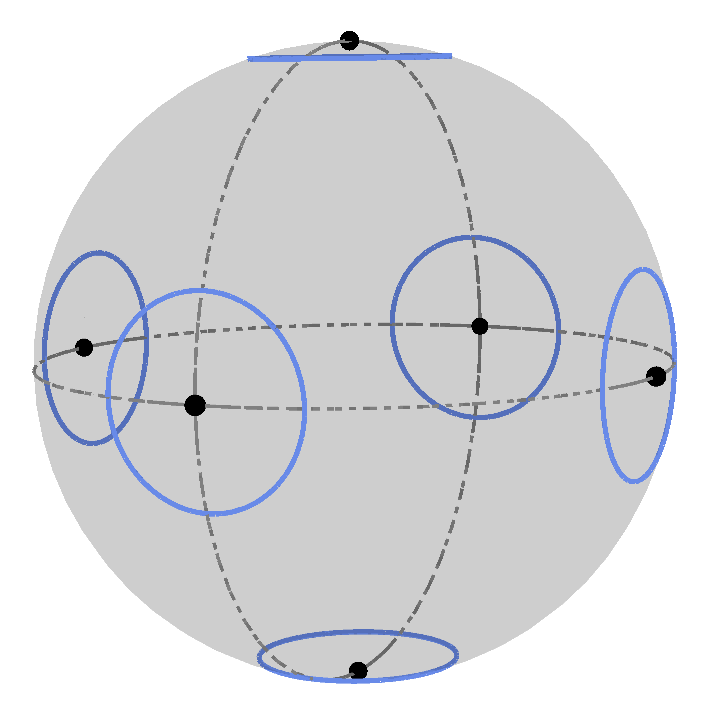}
        \caption{\centering{$\lambda=0$.}}
    \end{subfigure}
    \vspace{2mm}
    \begin{subfigure}[b]{0.32\textwidth}
        \includegraphics[width=\textwidth]{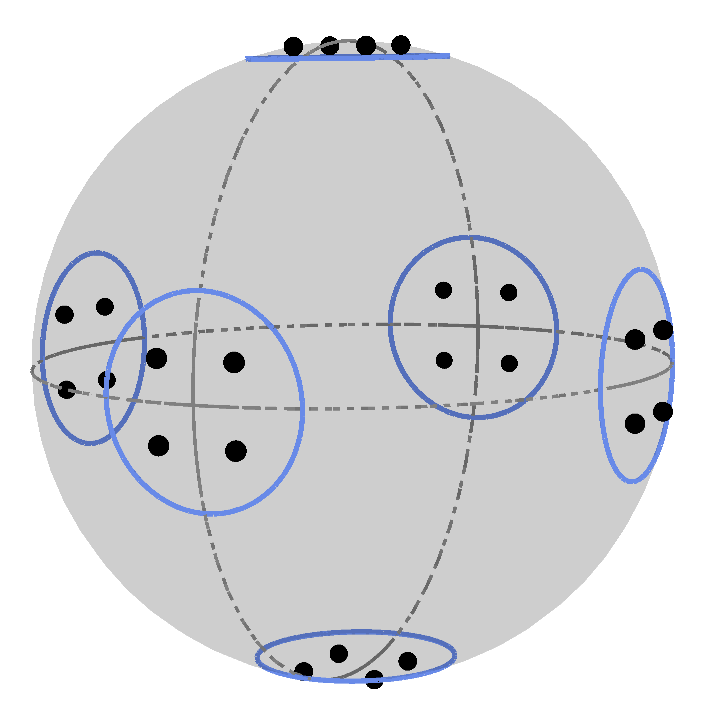}
           \caption{\centering{$\lambda \approx \lambda_{\rm crit} <0.5$.}}
    \end{subfigure}
    \vspace{2mm}
    \begin{subfigure}[b]{0.32\textwidth}
        \includegraphics[width=\textwidth]{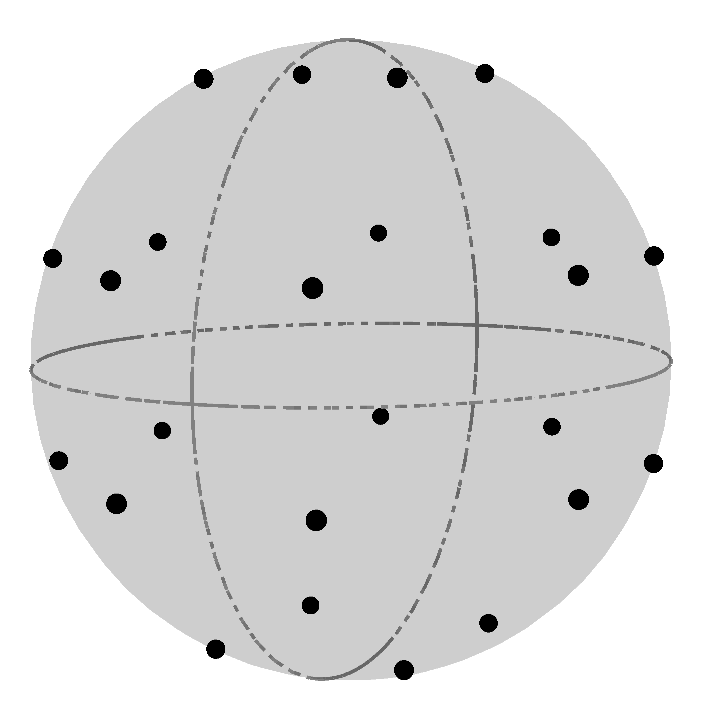}
           \caption{\centering{$\lambda=0.5$.}}
    \end{subfigure}
    \vspace{2mm}
       \begin{subfigure}[b]{0.32\textwidth}
        \includegraphics[width=\textwidth]{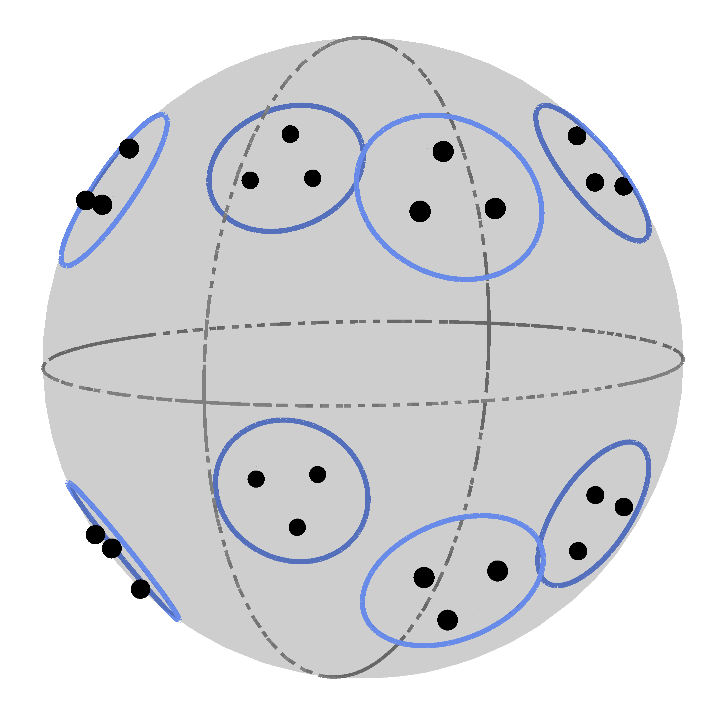}
        \caption{\centering{$\lambda \approx \lambda_{\rm crit} >0.5$.}}
    \end{subfigure}
    \vspace{2mm}
    \begin{subfigure}[b]{0.32\textwidth}
        \includegraphics[width=\textwidth]{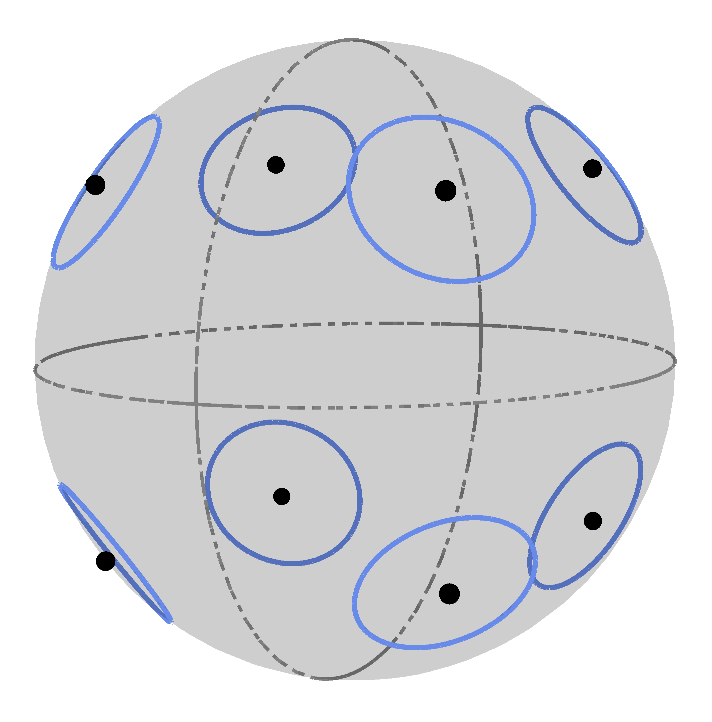}
           \caption{\centering{$\lambda=1$.}}
    \end{subfigure}
    \vspace{-5pt}
 \caption{A dimensionally-reduced depiction of the $8 \to 16$ mass case, where the masses move along trajectories parameterised by $\lambda$. For values of $\lambda < \lambda_{\rm crit}$, a shared horizon encompasses the black holes, which vanishes as they are moved further apart (see Section \ref{sec:horizons}).}
    \label{fig:spheres}
\end{figure}

\paragraph{8-Mass to 16-Mass Model:} The coordinates of the masses in the regular 8-mass model are given by all possible permutations of
\begin{equation} \label{8m}
   (w,x,y,z) = (\pm1,0,0,0) \, ,
\end{equation}
whilst those of its dual, the 16-mass model, consist of all permutations of
\begin{equation} \label{16m}
    (w,x,y,z) = \frac{1}{2}(\pm1,\pm1,\pm1,\pm1) \, .
\end{equation}
As an example, the mass at position $\frac{1}{2}(1,1,1,1)$ is split into 4 masses, each of which moves towards one of the following vertices of the 8-mass model: $(1,0,0,0)$, $(0,1,0,0)$, $(0,0,1,0)$ and $(0,0,0,1)$. Figure \ref{fig:spheres} shows a dimensionally-reduced version of this particular configuration, where one dimension is suppressed to show 6 masses on the surface of a 2-sphere at $\lambda=0$, instead of 8 masses on a 3-sphere. These are each split into 4 new masses and moved along trajectories on the surface of the 2-sphere, until they recombine at $\lambda=1$ to form a dimensionally-reduced version of the 16-mass model (8 masses on a 2-sphere). The blue rings highlight the apparent horizon around each of the original masses, which becomes a collective horizon when the masses are moved a small distance apart, and vanishes when the black holes become sufficiently separated (see Section \ref{sec:horizons} for further details).

\paragraph{24-Mass to 24-Mass Model:} The coordinates of the masses in the original 24-mass model are given as all possible permutations of
\begin{equation}
   (w,x,y,z) = \frac{1}{\sqrt{2}}(\pm1,\pm1,0,0) \, .
\end{equation}
The dual lattice in this case is also a 24-mass model, but with the locations of the final masses given by Eqs. (\ref{8m}) and (\ref{16m}).

\paragraph{120-Mass to 600-Mass Model:} The positions of the masses for the 120-mass model are also given by Eqs. (\ref{8m}) and (\ref{16m}), along with the additional 96 mass points located at all even permutations of
\begin{equation}
   (w,x,y,z) =   \frac{1}{2} (\pm \varphi, \pm 1, \pm \varphi^{-1}, 0) \, ,
\end{equation}
where $\varphi = (1+\sqrt{5})/2 = 1.618 \dots$ is the golden ratio. The dual of this lattice is the 600-mass model, with masses located at all permutations of
\begin{subequations}
\begin{align}
    &\frac{1}{\sqrt{8}}(0, 0, \pm 2, \pm 2)\\
    &\frac{1}{\sqrt{8}}(\pm 1, \pm 1, \pm 1, \pm \sqrt{5})\\
    &\frac{1}{\sqrt{8}}(\pm \varphi^{-2}, \pm \varphi, \pm \varphi, \pm \varphi)\\
    &\frac{1}{\sqrt{8}}(\pm \varphi^{-1}, \pm \varphi^{-1}, \pm \varphi^{-1}, \pm \varphi^2) \, ,
\end{align}
\end{subequations}
and all even permutations of
\begin{subequations}
\begin{align}
   &\frac{1}{\sqrt{8}} (0, \pm \varphi^{-2}, \pm 1, \pm \varphi^2)\\
   &\frac{1}{\sqrt{8}} (0, \pm \varphi^{-1}, \pm \varphi, \pm \sqrt{5})\\
   &\frac{1}{\sqrt{8}} (\pm \varphi^{-1}, \pm 1, \pm \varphi, \pm 2) \, .
\end{align}
\end{subequations}
The positions of all of these lattice and dual-lattice vertices have been verified by evaluating the Kretschmann scalar $\widehat{K}= R_{abcd}^{(3)}R^{(3) abcd}$,
where $R^{(3) abcd}$ is the Riemann tensor for the 3-space, at each of the original positions of the masses and vertices.

\subsection{Trajectories of the new black holes}
\label{sec:traj}

Let us now be more precise about the trajectories discussed above, and that were illustrated with examples in Figs. \ref{fig:cube} and \ref{fig:spheres}. Our basic aim is to be able to consider instants of time at which the black holes will be positioned at different points along the curves that connect the centres of our lattice cells to their vertices. We have euphemistically referred to this as ``moving'' the black holes along these trajectories, but the reader should be aware that what we are really doing is comparing different configurations of black hole positions in different universes, each at their maximum of expansion. We consider this to be a type of quasi-static approach to structure formation, where the dynamics themselves are neglected, but the consequences of clustering in these instantaneously static configurations can be explored in detail  (akin to the study in Ref. \cite{cadez}, where instantaneously static black holes at different spatial separations in an asymptotically flat space were compared).

The trajectories we are considering are examples of the locally rotationally symmetric curves identified in Ref. \cite{evolution3}. At least two reflection symmetric planes will always intersect along each of these paths \cite{silence}, and the property than any reflection symmetric surface is totally geodesic \cite{Eisenhart} means that these trajectories must themselves be geodesics in the geometry of the initial 3-space. This follows for both the full geometry, and the conformal geometry in Eq. (\ref{eq:conformal}). In terms of parameterising these trajectories, it is more straightforward to consider them in the round geometry that constitutes the conformal space. We therefore require $\ddot{x}^{\mu} + \tilde{\Gamma}^{\mu}_{\rho\sigma}\dot{x}^{\rho}\dot{x}^{\sigma}=0$ along each trajectory, where $\tilde{\Gamma}^{\mu}_{\rho\sigma}$ are the Christoffel symbols for the round metric of a 3-sphere, and over-dots are derivatives with respect to an affine parameter.

A simple way to determine the position at any point along our chosen trajectories is to label the position 4-vector in the Euclidean embedding space at any point along the path as $v$, and to parameterise the trajectory itself by $\lambda$, such that
\begin{equation}
\label{eq:path}
v(\lambda) \equiv \left( w(\lambda),x(\lambda),y(\lambda),z(\lambda) \right) = (1-\lambda) v_{\rm original} + \lambda v_{\rm dual} \, ,
\end{equation}
where $v_{\rm original}$ is the original position of one of the masses in the regular lattice, and  $v_{\rm dual}$ is the position of one the masses in the dual lattice (or, equivalently, one of the vertices in the original lattice). Every point along the trajectory is then given by a unique value of $\lambda$ $\in$ $[0,1]$. This choice of parameterisation has been made so that at $\lambda = 0$ the new masses have not been moved from the original position of the black holes in the original lattice, and so that at $\lambda=1$ they have been moved the maximal amount (and are therefore at the positions of one of the masses in the dual lattice). Thus, the analysis of regular lattice configuration performed in Ref. \cite{tim} becomes the limiting case, when $\lambda = 0$ or $1$.

The reader will note that the paths described by Eq. (\ref{eq:path}) cut through the embedding space $E^4$ in a straight line, and therefore represent a chord of the conformal 3-sphere. We must therefore impose a further constraint in order for the masses to move along the surface of the 3-sphere, along geodesic arcs. To do this, it is useful to transform from Cartesian to hyperspherical coordinate systems, so that we can write the trajectory parameterised in Eq. (\ref{eq:path}) as
\begin{equation}
    v(\lambda) = (r\cos{\rchi}, r\sin{\rchi}\cos{\theta}, r\sin{\rchi}\sin{\theta}\cos{\phi},  r\sin{\rchi}\sin{\theta}\sin{\phi}) \, ,
\end{equation}
where $r, \rchi, \theta$ and $\phi$ should here all be considered as functions of $\lambda$. Now, a mass on the chord described by the path $v(\lambda)$ has the same angular coordinates as one on the corresponding arc, which we choose to write as $v'(\lambda)$. Recognising the 3-sphere is of unit size (with $r=1$) therefore gives
\begin{equation}
\label{eq:newpath}
    v'(\lambda) = (\cos{\rchi}, \sin{\rchi}\cos{\theta}, \sin{\rchi}\sin{\theta}\cos{\phi},  \sin{\rchi}\sin{\theta}\sin{\phi}),
\end{equation}
where $\rchi=\rchi(\lambda), \theta= \theta(\lambda)$ and $\phi=\phi(\lambda)$ are obtained from transforming the $(w,x,y,z)$ coordinate positions from Eq. \eqref{eq:path} into hyperspherical coordinates. As long as we know $v_{\rm original}$ and $v_{\rm dual}$, we can therefore locate any point along any of the trajectories  along which we want to move our new black holes. These are exactly the two quantities we gave for each of the six possible regular lattice configurations in Section \ref{sec:ext}, above.

As an example, let us consider the $5$-mass to $5$-mass model, with $v_{\rm original}$ and $v_{\rm dual}$ as given in \eqref{eq:orig} and below. If we choose the particular trajectory that starts at $v_{\rm original} = (1,0,0,0)$, and ends at $v_{\rm dual}= \sqrt{\frac{5}{16}} \left( \frac{1}{\sqrt{5}},-1,-1,-1 \right)$, then Eq. \eqref{eq:path} gives
\begin{eqnarray}
v(\lambda) &=& (1-\lambda) (1,0,0,0) + \lambda \sqrt{\frac{5}{16}}\left(\frac{1}{\sqrt{5}},-1,-1,-1\right) \\[5pt]
&=& \left(1-\frac{3\lambda}{4}, -\lambda\sqrt{\frac{5}{16}}, -\lambda\sqrt{\frac{5}{16}}, -\lambda\sqrt{\frac{5}{16}}\right) \, .
\end{eqnarray}
Using the standard transformations from Cartesian to hyperspherical coordinate systems, given in Eqs. (\ref{eq:transformations}), we then obtain the functions
\begin{subequations}\label{eq:angles}
\begin{align}
\label{eq:angles:1}
\rchi(\lambda) & = \arccos\left(1-\frac{3\lambda}{4}\right),
\\
\label{eq:angles:2}
\theta(\lambda) & = \arccos\left(\frac{-\lambda \sqrt{\frac{5}{16}}}{\sin{\rchi(\lambda)}}\right),
\\
\label{eq:angles:3}
\phi(\lambda) & = \arccos\left(\frac{-\lambda \sqrt{\frac{5}{16}}}{\sin{\rchi(\lambda)}\sin{\theta(\lambda)}}\right),
\end{align}
\end{subequations}
which can be substituted into the expression \eqref{eq:newpath} to obtain the full path for the mass in the conformal 3-space. Of course, Eqs. \eqref{eq:angles:1}, \eqref{eq:angles:2} and \eqref{eq:angles:3} only describe one of the 4 paths we wish to consider, within one of the 5 cells in the model. There are altogether 3 angular functions for each of the 4 vertices of each of the 5 lattice cells - a total of 60 functions, describing 20 trajectories. For the largest $n$ model, where the lattice and dual lattice have 120 and 600 masses, there are a total of 7200 functions describing 2400 trajectories. These trajectories have been checked to ensure they are geodesics of the conformal 3-sphere, as required from the discussion at the beginning of this section.

\section{Apparent Horizons and Shared Horizons}
\label{sec:horizons}

We now have a method to place mass points at arbitrary positions along the curves that connect the centres and vertices of lattice cells. If we choose to split each of the black holes in our original, regular lattice configurations into $C_n$ new objects that each have the same mass, and then move all of them away from the centre of every cell at the same rate, then we have the basis for a cosmological model that simultaneously allows for clustering on small scales, while maintaining a degree of statistical homogeneity and isotropy on large scales. It remains to demonstrate that the new objects in question really are black holes, and to evaluate both the masses of these black holes, the masses of the clusters that they form, and the positions of their horizons. In this section we consider the problem of identifying the positions of apparent horizons within our exact initial data. This is sufficient to demonstrate the existence of black holes and a cosmological region. In the next section we will move on to measures of mass, which is a somewhat more complicated issue.

The positions of black hole horizons is particularly important in the models we will be discussing, as the possible merging or sharing of horizons could change the number of black holes that an observer in the cosmological region of the space infers. Indeed, it could even impact upon the existence of a cosmological region at all. For our models to represent cosmological models we require, as a necessary condition, that the horizons of neighbouring black holes should remain distinct, and not overlap. We will call a region of space ``cosmological'' if it is entirely bounded by apparent horizons, and is not in causal contact with any asymptotically flat region of space. Any time-like observer in such a cosmological region will then be able to directly infer the presence of black holes, through the influence of their gravitational field, but will not be able to relocate themselves to a region that is arbitrarily far from all of them. The number of black holes in this cosmological region, as determined by the observer, will be equal to the number of distinct, closed, marginally outer trapped surfaces that bound it.

An interesting complication that arises, when considering the clustering of black holes, is that if these objects are brought close enough together then an extra apparent horizon can appear, which encompasses them both. The black holes retain their own individual horizons, but also acquire this new shared horizon. In such a circumstance, there may be multiple mass points that correspond to a single black hole (as inferred by an observer in a cosmological region). An analogous phenomenon is already well known for two black holes sufficiently close enough to each other in an asymptotically-flat space \cite{brillandlind,cadez}, but brings some new twists in the cosmological context. We will discuss this further below, as well as calculate precisely when such shared horizons should be expected to appear. First of all though, we will outline our procedures for locating apparent horizons.

\subsection{Locating apparent horizons}

Apparent horizons are defined as the outermost marginally outer trapped surfaces that exist within a space. In 3-dimensional space these are 2-dimensional closed surfaces, and are often used in demarking the edge of a black hole, as a proxy for the event horizon. The condition for a surface to be marginally outer trapped is mathematically expressed as the condition that the expansion of the outward-pointing null normal to the surface should have vanishing expansion, so that $\nabla_\mu k^{\mu} =0$. The precise location of apparent horizons can be found numerically \cite{shib}, or located using approximate or analytic methods. Here we will use two different techniques which contain complementary information about the properties of the apparent horizon, and that can be used together to give a measure of asphericity in its shape. We will refer to these as {\it the area method} and {\it the Weyl tensor method}, and they proceed as follows:

\paragraph{The Area Method:} In the time-symmetric case, all marginally outer trapped surfaces correspond to closed extremal surfaces \cite{gib}. This follows immediately from the fact that the outward pointing null normal can be decomposed into a  time-like vector $u^{\mu}$ and a space-like vector $e_1^{\mu}$, such that
\begin{equation} \label{exhor}
\nabla_{\mu} k^{\mu} = \frac{1}{\sqrt{2}} \nabla_{\mu} ( u^{\mu} + e_1^{\mu}) = \frac{1}{\sqrt{2}} \nabla_{\mu} u^{\mu}+\frac{1}{\sqrt{2}} \nabla_{\mu} e_1^{\mu} = 0 \, .
\end{equation}
If the $e_1^{\mu}$ exists within the initial hypersurface, then $u^{\mu}$ is orthogonal to it. From the vanishing of every component of the extrinsic curvature of the initial data, required for time-reversal symmetry, we then have $\nabla_{\mu} u^{\mu}=0$ (as this is proportional to $K$). This leaves us with the condition that $\nabla_{\mu} e_1^{\mu} =0$ at the location of any apparent horizon, if $e_1^{\mu}$ is orthogonal to that horizon. Any closed surface that has a normal vector that obeys this condition is automatically an extremal surface in the 3-space \cite{Eisenhart}. In our case, the apparent horizons of black holes will all correspond to minimal surfaces.

If the gravitational field in the vicinity of each black hole is close to spherically symmetric, then one way of estimating the position of the apparent horizon is to approximate it as being a sphere in the conformal geometry, with constant radial coordinate $\rchi=\rchi_h$. The position of this sphere can be found by considering concentric shells with different values of $\rchi$, and then determining the value of $\rchi$ that gives a sphere with the minimum area, as measured by evaluating 
\begin{equation}
\label{eq:area}
A(\rchi) = \int_0^{2\pi} \int_0^{\pi} \psi^4 \sin^2(\rchi) \sin(\theta) d\theta d\phi \, .
\end{equation}
The benefit of this method is that it is relatively simple, and that it gives us an approximation for the position of the apparent horizon in every direction away from the centre of the black hole. The drawback is that it should only be expected to be accurate when the black hole horizon is close to spherical (i.e. it will become increasingly inaccurate as we start to consider situations in which the horizon is increasingly distorted). This is not necessarily a problem if the black holes are all well separated, but could lead to problems if the black holes are very close together.

\paragraph{The Weyl Tensor Method:} To more accurately determine the position of horizons in our initial data we can further examine the content of the constraint equations. In particular, if we note that the initial data is extrinsically flat then we can use the Ricci identities and the Gauss embedding equation to write \cite{evolution3}
\begin{equation} \label{ER}
E_{\mu\nu} =  \mathcal{R}_{\mu\nu} \, ,
\end{equation}
where $E_{\mu \nu}$ is the electric part of the Weyl tensor with respect to the 4-velocity $u^{\mu}$, and $\mathcal{R}_{\mu\nu}$ is the Ricci tensor of our 3-dimensional hypersurface. This result is useful as we can calculate $\mathcal{R}_{\mu\nu}$ explicitly using the results from Section \ref{sec:initial}, and because the Bianchi identities expressed in an orthonormal tetrad allow us to write \cite{evolution3}
\begin{equation}
 \mathbf{e}_1(E^{11}) = 3 a_1 E^{11} + n_{23}(E^3_{\,3} - E^2_{\,2}) \, ,
\end{equation}
where $\mathbf{e}_1$ is a frame derivative in a direction normal to the apparent horizon, and subscripts $1$, $2$ and $3$ in this equation correspond to frame components. The quantities $a_1$ and $n_{23}$ are the rotation coefficients of the spatial frame \cite{orth}, and in writing this equation we have used the result that the normal to the apparent horizon is a principal Ricci direction of the initial 3-space, and hence also a principal direction of $E_{\mu \nu}$ \cite{evolution3}.

At the location of an apparent horizon we know from Eq. (\ref{exhor}) that $\nabla_{\mu} e_1^{\mu}=0$, which in turn implies that $a_1 =0$. If we can now locate positions on the apparent horizon where $E^3_{\,3} = E^2_{\,2}$ then we are left with the simple condition $\mathbf{e}_1(E^{11}) =0$. The locally rotationally symmetric curves studied in Ref. \cite{silence} obey exactly this property, which means that if we arrange our coordinate system so that $\mathbf{e}_1 = \psi^{-2} \partial_{\rchi}$ points along these curves then a necessary condition for determining the location of an apparent horizon is
\begin{equation} \label{eq:weyl}
\mathbf{e}_1 (E^{11}) =  0
\qquad {\rm or, equivalently,} \qquad
\frac{1}{\psi^2}\frac{\partial}{\partial \rchi} (\psi^{-4} {\mathcal R}_{\rchi \rchi}) = 0 \, .
\end{equation}
In other words, if we plot $E^{11}$ or $\psi^{-4} {\mathcal R}_{\rchi \rchi}$ along a locally rotationally symmetric curve parameterised by $\rchi$, then marginally outer trapped surfaces will be located at the points at which this function is extremised. This method is expected to more precise than the area method outlined above, especially when horizons are aspherical, but is only sufficient to locate the points at which the horizon intersects locally rotationally symmetric curves. For further details of the Weyl tensor method the reader is referred to Ref. \cite{evolution3}.

\subsection{Critical values of $\lambda$}

\begin{figure}[b!]
    \centering
  \includegraphics[width=0.85\textwidth]{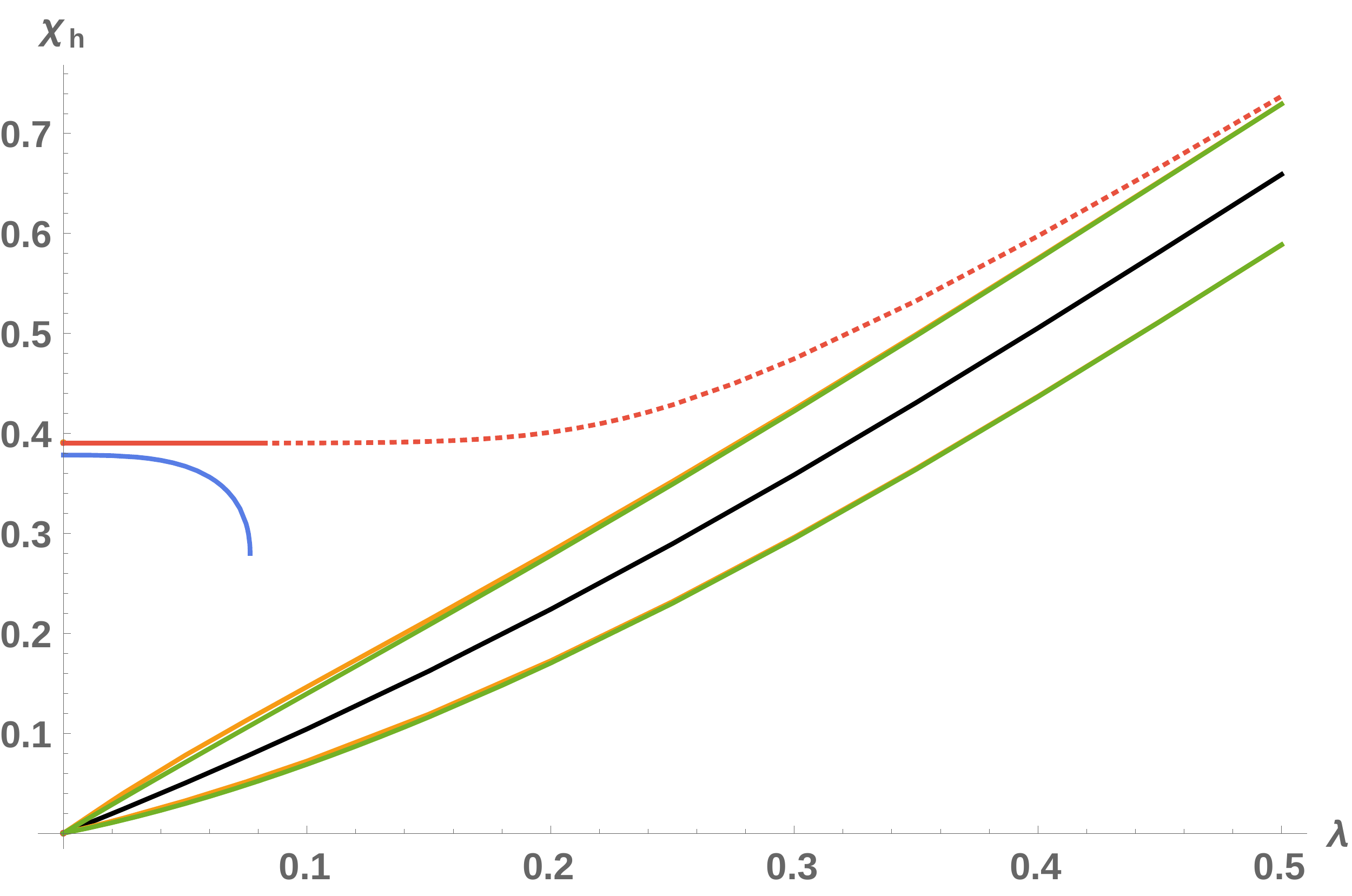}
     \caption{\label{fig:5horizons}The values of $\rchi_h$ for both individual and shared horizons at different values of $\lambda$ for the 5-mass case. The blue and red lines are the collective horizons from the Weyl and area methods, respectively. The dashed line indicates that this horizon only exists for values of $\lambda$ up to $\lambda_{\rm crit}$, but can always be found using the area method. We plot the position of one of the black holes as it moves radially outward from $\rchi=0$, shown by the black line. The orange and green lines are largely indistinguishable, and show the individual horizon either side of this black hole, calculated using the Weyl and area method respectively.}
     \vspace{20pt}
\end{figure}

Let us now consider the results of applying these two methods to locate the horizons that exist when we cluster masses, in the manner explained in Section \ref{sec:structure}. First of all let us consider the 5-mass lattice, with each of the original masses split into 4 new black holes. When we increase the parameter $\lambda$ away from zero we find that there are, in fact, multiple horizons. Each of the 20 new black holes is contained within its own individual horizon. These horizons exist for all values of $\lambda >0$. As well as these individual horizons, however, there can also be seen to be horizons that encompass multiple black holes. These shared horizons only exist when $\lambda$ is below some critical value, i.e. when $\lambda < \lambda_{\rm crit}$. When they do exist, they encompass all of the 4 masses that cluster around the original positions of the mass points, as shown in the dimensionally-reduced illustrations in Fig. \ref{fig:spheres}.

The precise positions of the apparent horizons for the 20 black holes derived from the 5-mass model are shown in Fig. \ref{fig:5horizons}. The individual horizons of each of the 20 black holes are shown as the orange and green lines in this plot, and corresponds to the coordinate radius of each of the horizons as determined using the Weyl tensor method and the area method, respectively. They are largely indistinguishable by eye, which demonstrates that these horizons are always close to spherical (otherwise the area method would be inaccurate, and differ from the Weyl tensor method). This makes sense, as the shape of an apparent horizon around just one black hole should not be expected to be strongly distorted by the presence of others, which are in fact a large proper distance away. The remaining two lines in this plot show the position of the shared horizon, as determined by our two different methods. These lines are clearly distinguishable from each other, which indicates that the shared horizon for the cluster of 4 black holes is not entirely spherical. In determining the position of the shared horizon we rotate the lattice so that $\rchi=0$ at the centre of the cell, making the displayed value of $\chi$ as close as possible to something resembling the radius of the horizon.

\begin{figure}[tbp]
    \centering
    \begin{subfigure}[b]{0.42\textwidth}
        \includegraphics[width=\textwidth]{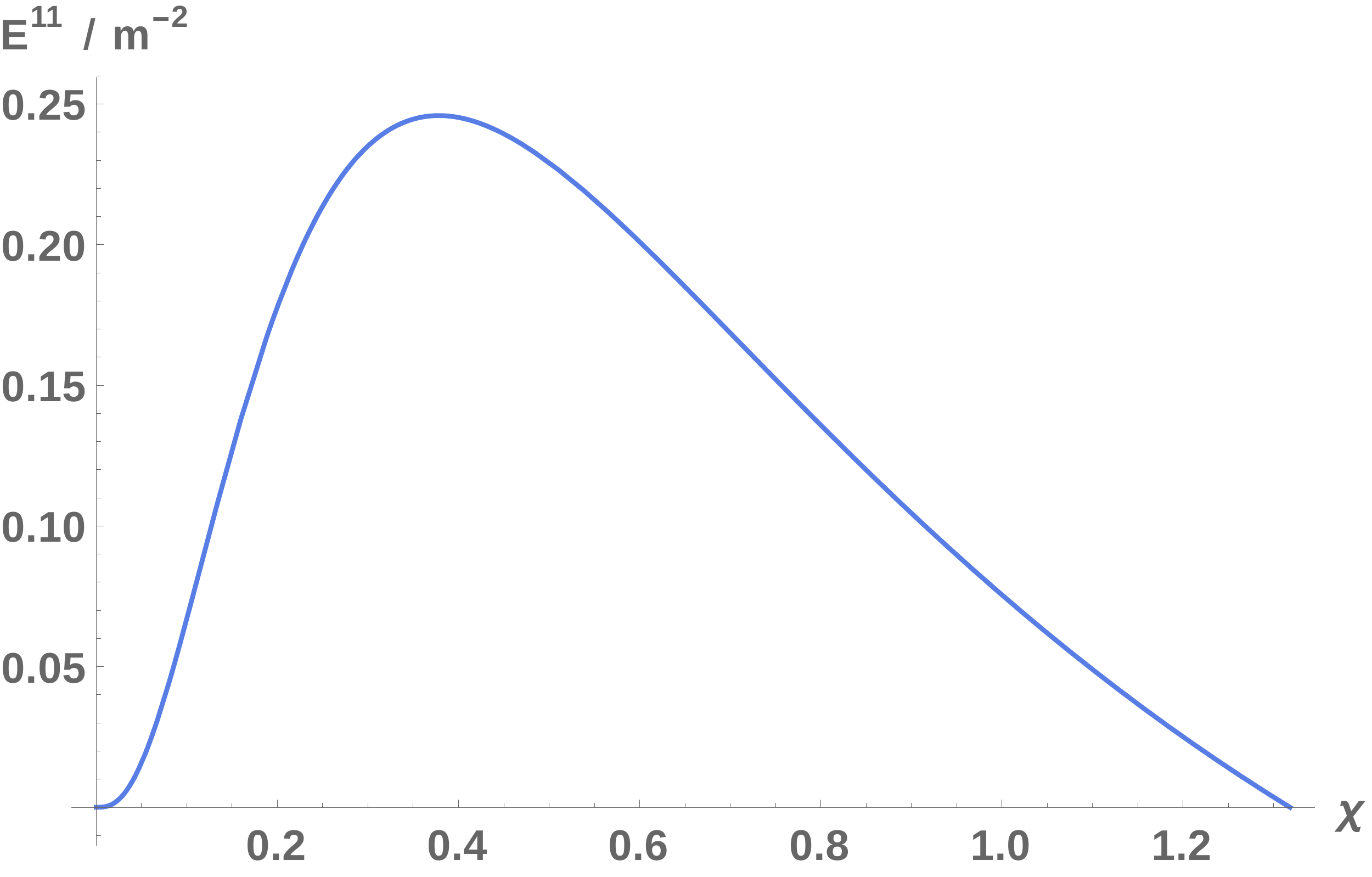}
        \caption{\centering{$\lambda=0$.}}
    \end{subfigure}
    \vspace{2mm}
    \begin{subfigure}[b]{0.42\textwidth}
        \includegraphics[width=\textwidth]{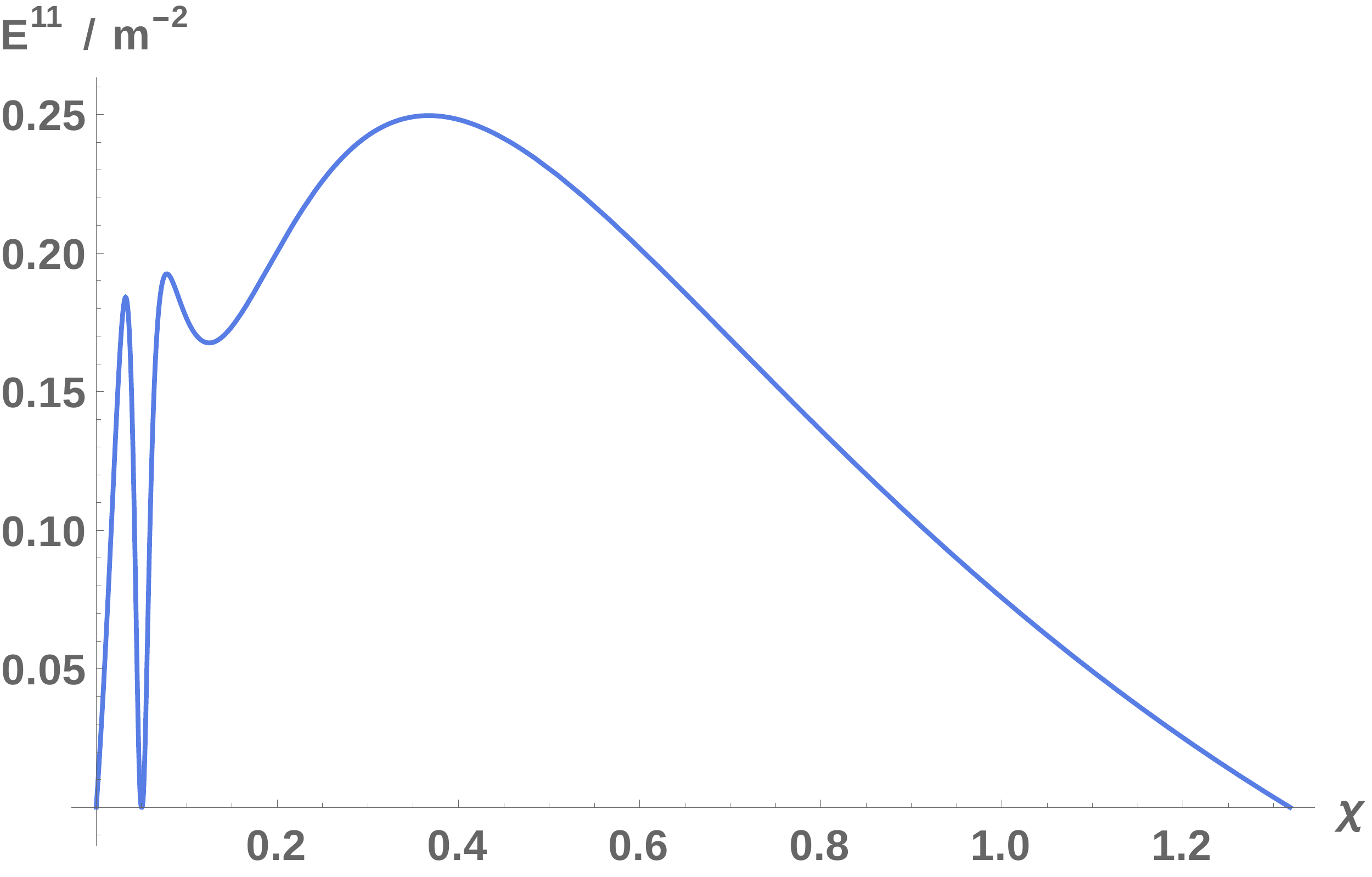}
           \caption{\centering{$\lambda=0.05$.}}
    \end{subfigure}
    \begin{subfigure}[b]{0.42\textwidth}
        \includegraphics[width=\textwidth]{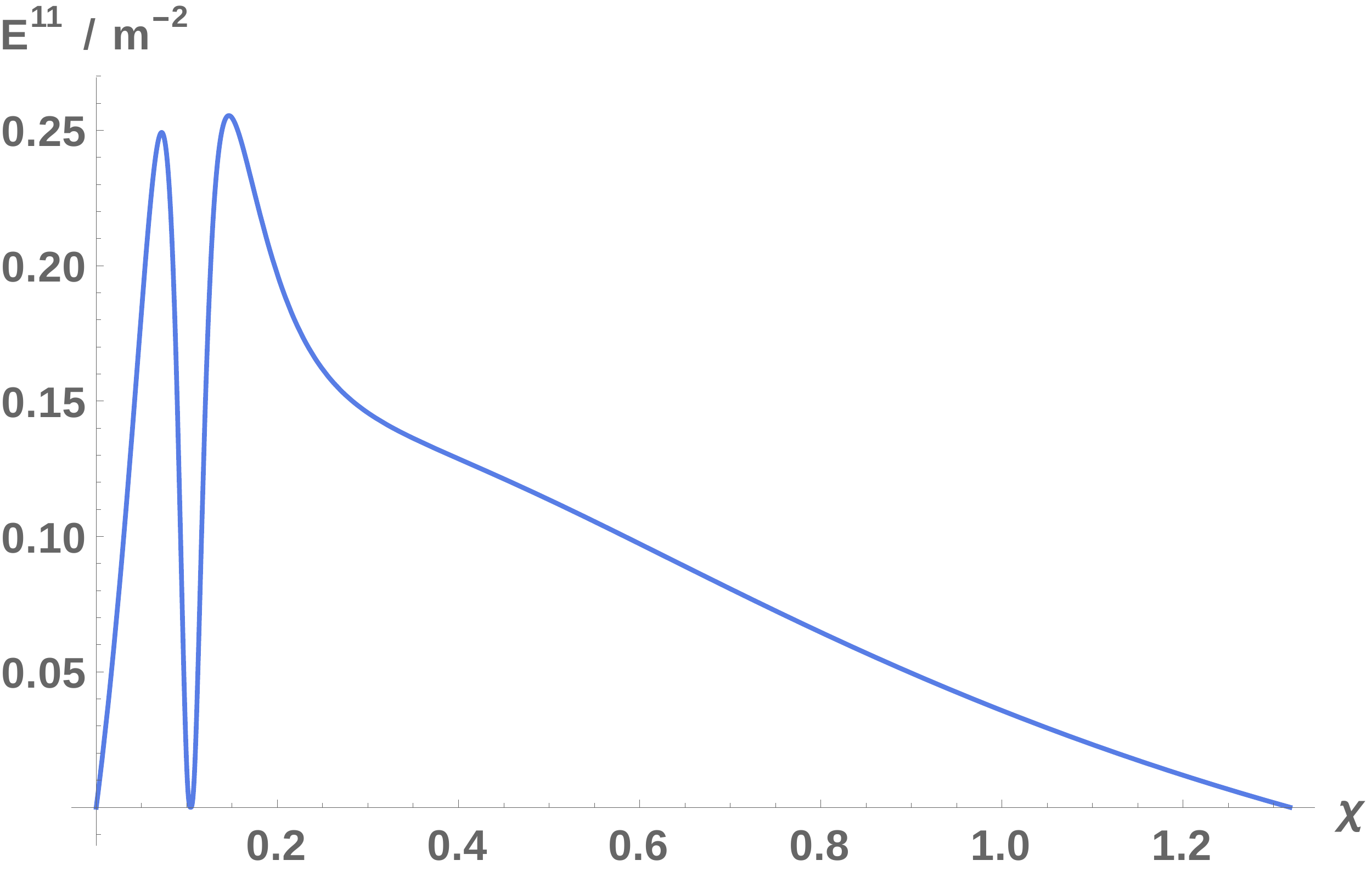}
           \caption{\centering{$\lambda=0.1$.}}
    \end{subfigure}
    \vspace{2mm}
    \begin{subfigure}[b]{0.42\textwidth}
        \includegraphics[width=\textwidth]{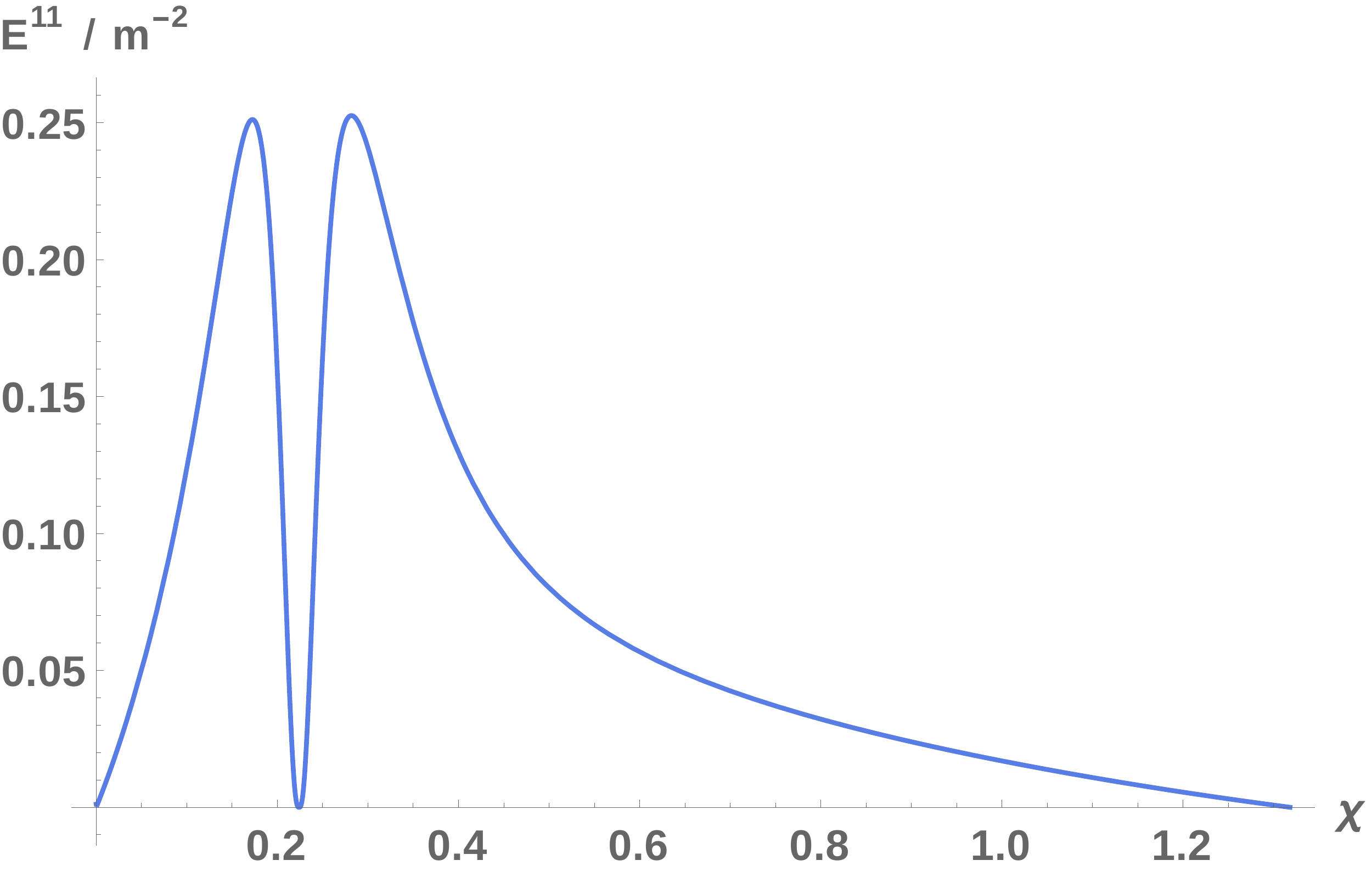}
           \caption{\centering{$\lambda=0.2$.}}
    \end{subfigure}
    \begin{subfigure}[b]{0.42\textwidth}
        \includegraphics[width=\textwidth]{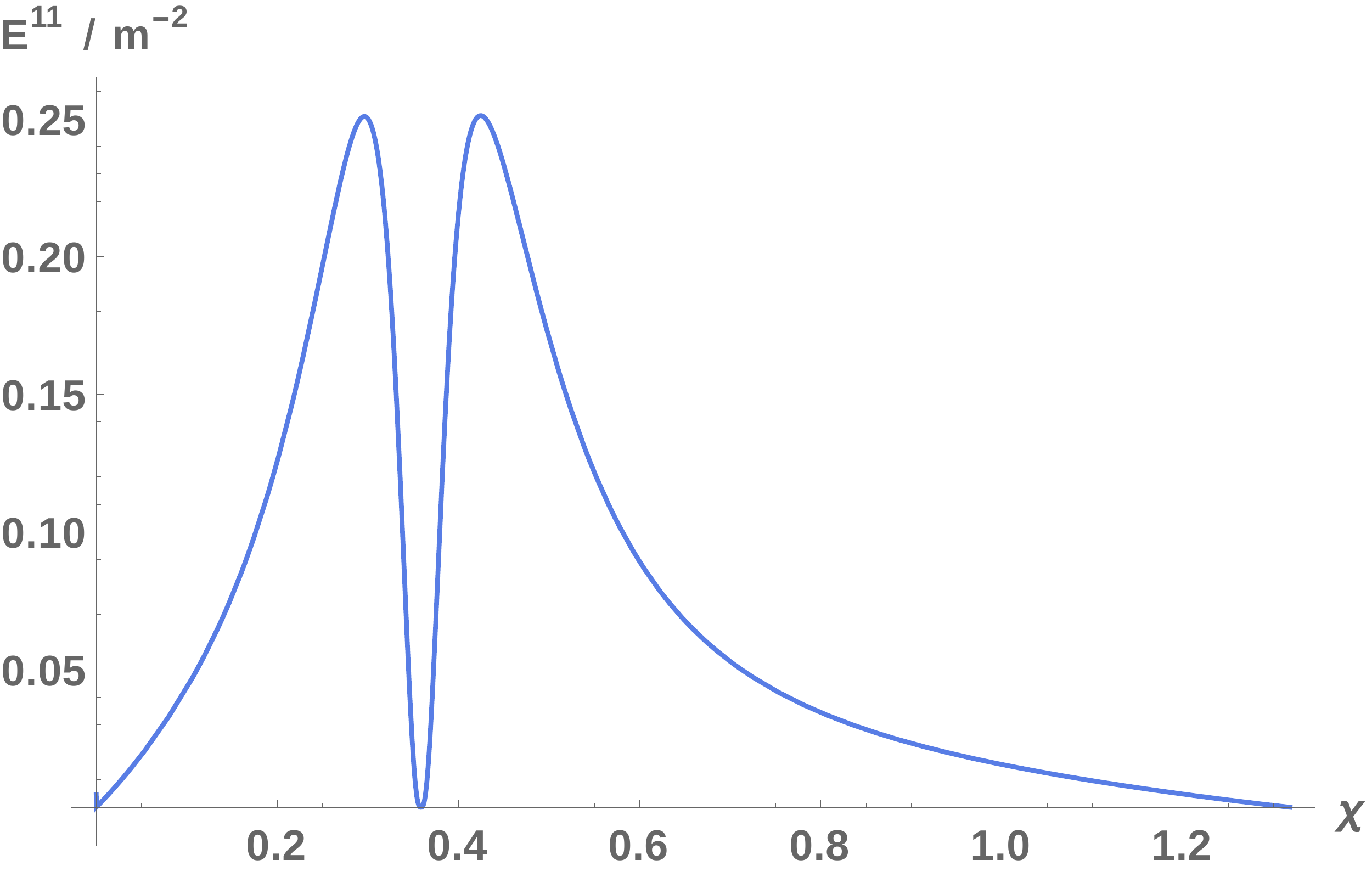}
           \caption{\centering{$\lambda=0.3$.}}
    \end{subfigure}
      \begin{subfigure}[b]{0.42\textwidth}
        \includegraphics[width=\textwidth]{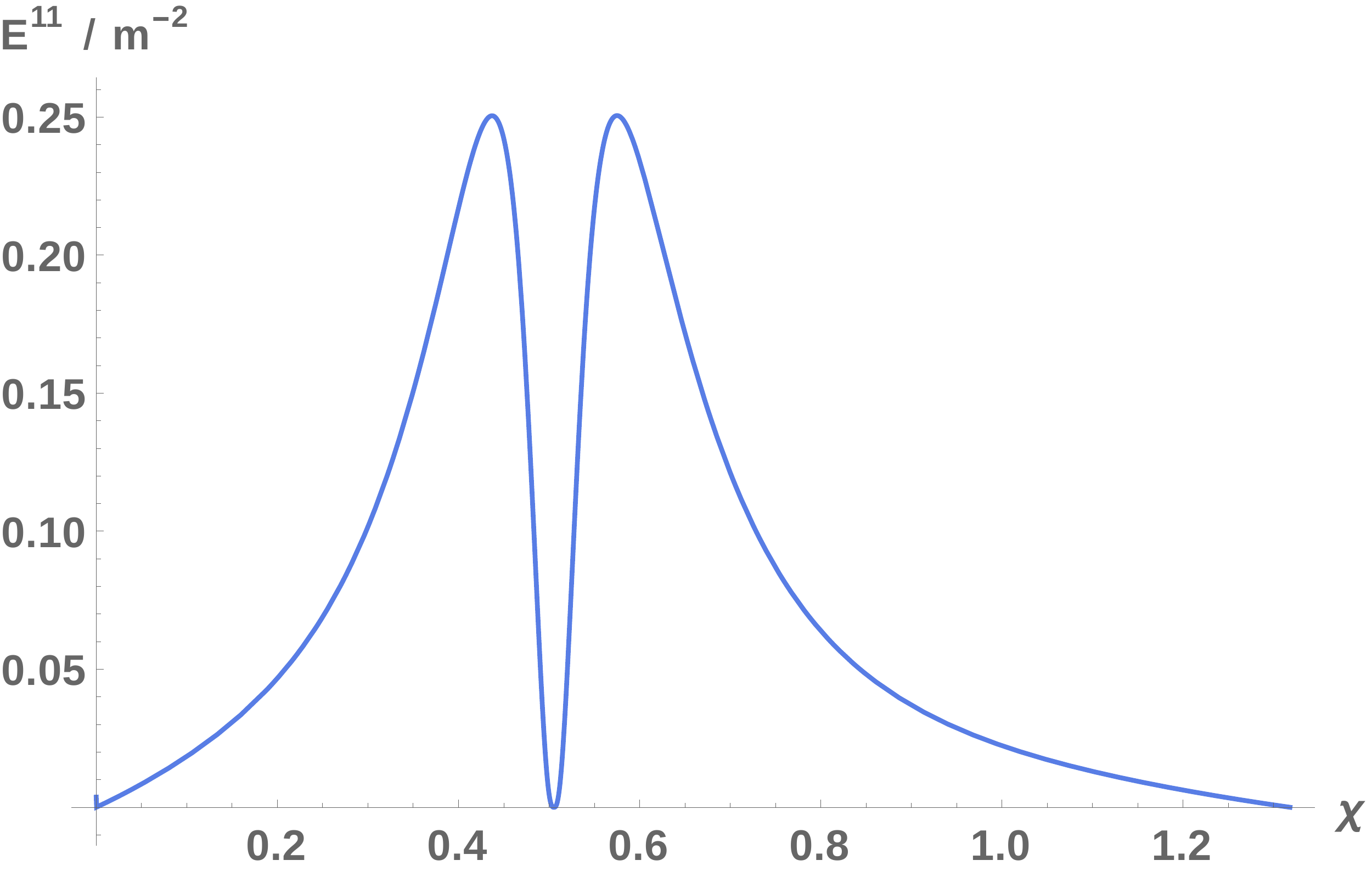}
           \caption{\centering{$\lambda=0.4$.}}
            \end{subfigure}
    \begin{subfigure}[b]{0.42\textwidth}
     \vspace{4mm}
        \includegraphics[width=\textwidth]{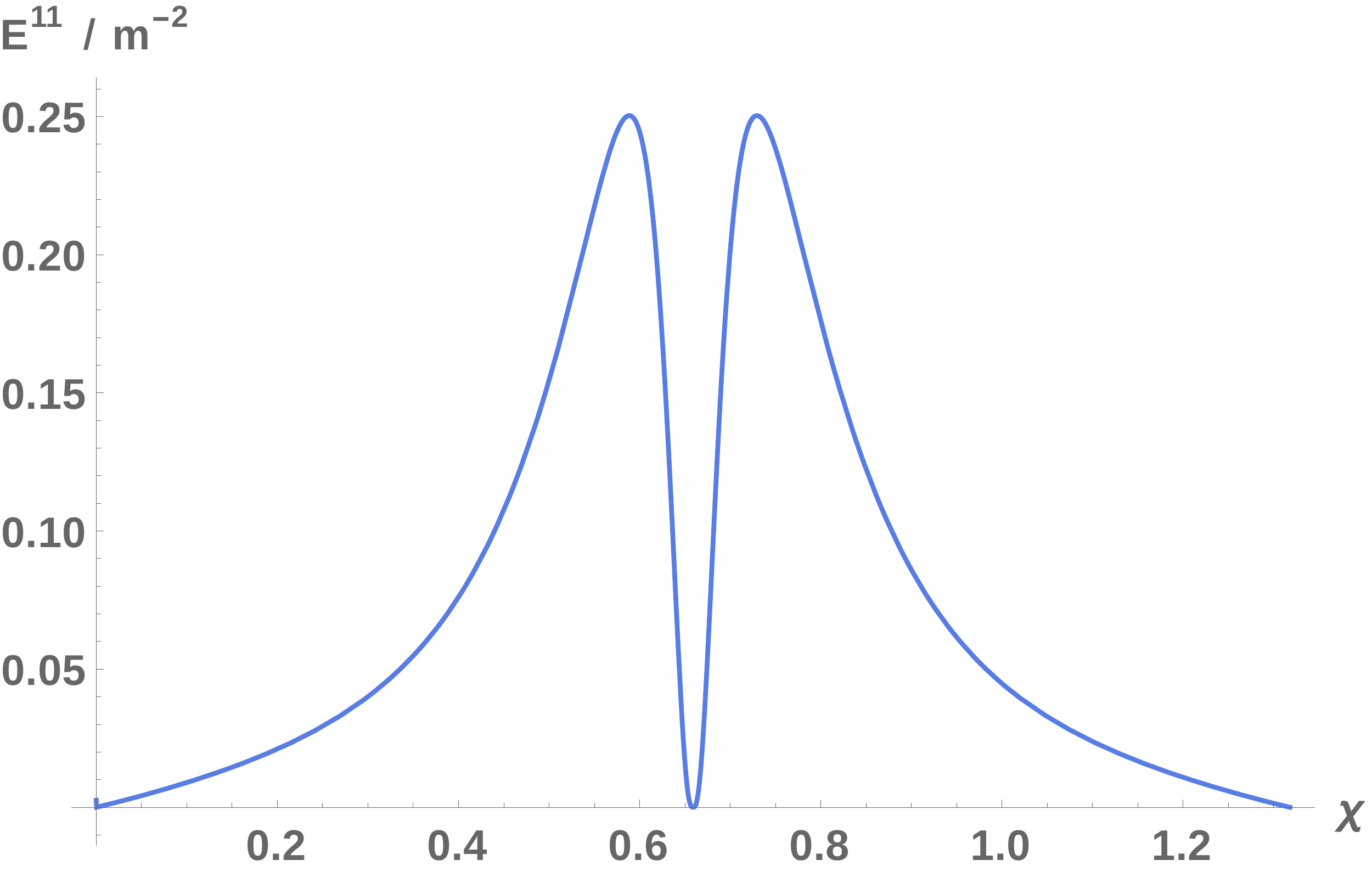}
           \caption{\centering{$\lambda=0.5$.}}
    \end{subfigure}
    \caption{The value of $E^{11}$ along locally rotationally symmetric curves, for different values of $\lambda$ in the 5-mass lattice.  $E^{11}$ is presented here in units of $m^{-2}$, where $m$ is the proper mass of each of the black holes (discussed in detail in Section \ref{sec:mass}).}
    \label{fig:5weyl}
\end{figure}

\begin{figure}[t!]
    \centering
    \vspace{20pt}
  \includegraphics[width=0.85\textwidth]{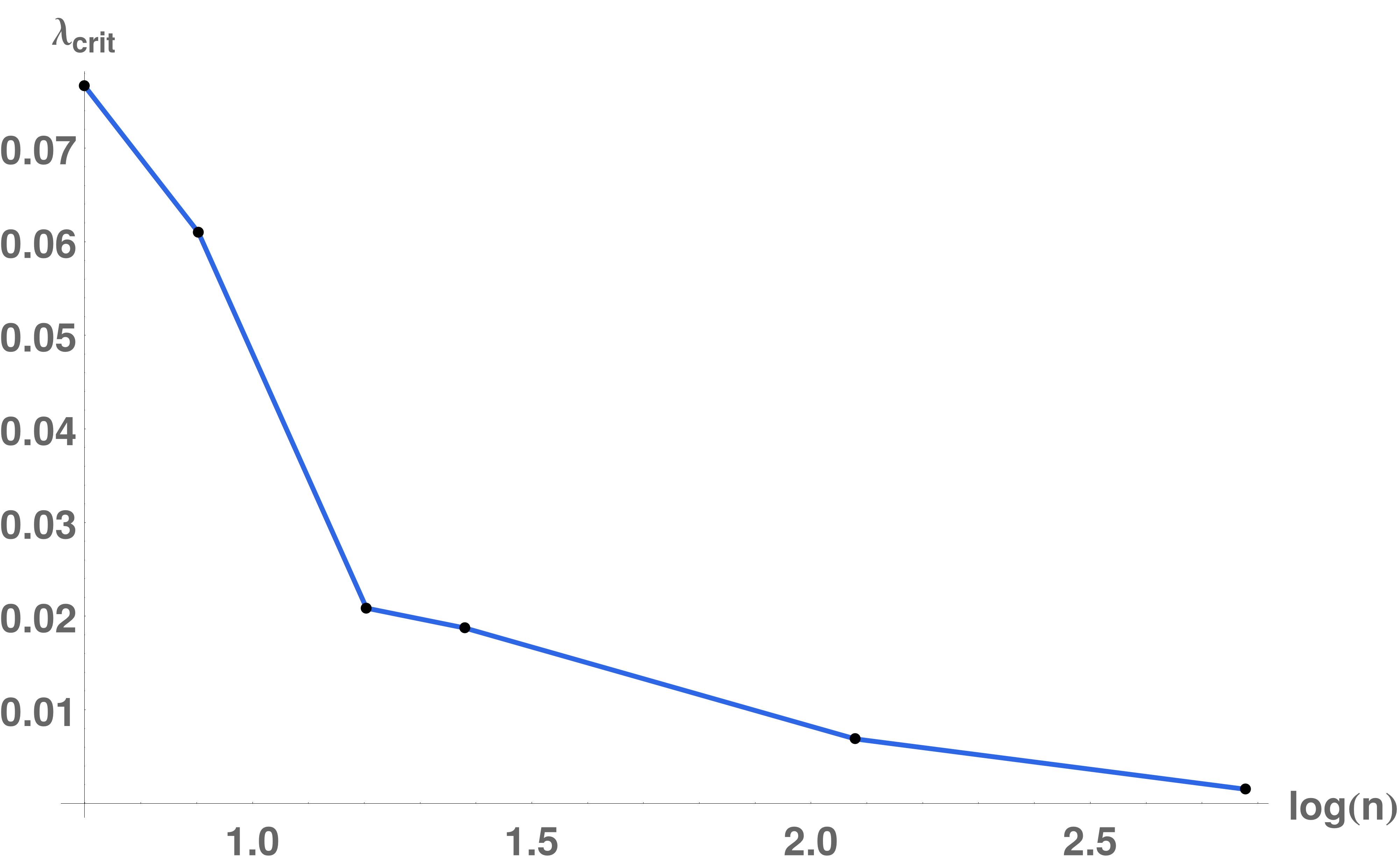}
     \caption{\label{fig:fraction}The critical value of $\lambda$, after which no shared horizons exist, as a function of the number of masses in the original lattice models.}
\end{figure}

The difference between the area and Weyl methods, when applied to the shared horizon, shows a number of other interesting features. Whilst the area method gives us a decent approximation to the location of the shared apparent horizon for small $\lambda$, it is not able to determine when such a collective horizon vanishes. This is because there is always a minimal surface of constant $\rchi$ that be can be found around any set of black holes. To find out the value of $\lambda_{\rm crit}$, beyond which the shared horizon vanishes, we must therefore use the electric part of the Weyl tensor. The value of $\lambda_{\rm crit}$ for the 5 black hole lattice can then be found by observing the point at which the blue line ends in Fig. \ref{fig:5horizons}, i.e. at $\lambda \simeq 0.077$. The area method still gives results beyond this limit, as indicated by the red dashed line, but these do not correspond to any actual trapped surfaces.

To further illustrate how it is the shared horizons that vanish, we plot the relevant frame component of the electric part of the Weyl tensor for the 5 mass case, for different values of $\lambda$, and along the curve that connects the cell centre with the vertex. These results are displayed in Fig. \ref{fig:5weyl} for $\lambda=0$, $0.05$, $0.1$, $0.2$, $0.3$, $0.4$ and $0.5$. To produce these plots we initially rotated the lattice so that a mass appeared at $\rchi=0$. Figure \ref{fig:5weyl} shows that there are multiple different points that satisfy Eq. \eqref{eq:weyl}, represented by the different maxima and minima along the curve. In each case we identify the maxima as apparent horizons. For $\lambda_{\rm crit} > \lambda >0$ there are three of these, the first two of which correspond to the individual black hole horizons around the black hole at $\rchi=0.05$, and the last of which corresponds to the shared horizon of the cluster. At $\lambda=0.05$ both types of horizon are very clear. However, at $\lambda =0.1$ the shared horizon has disappeared entirely.

\begin{figure}[tbp]
\vspace{0mm}
    \centering
    \begin{subfigure}[t]{0.49\textwidth}
        \includegraphics[width=\textwidth]{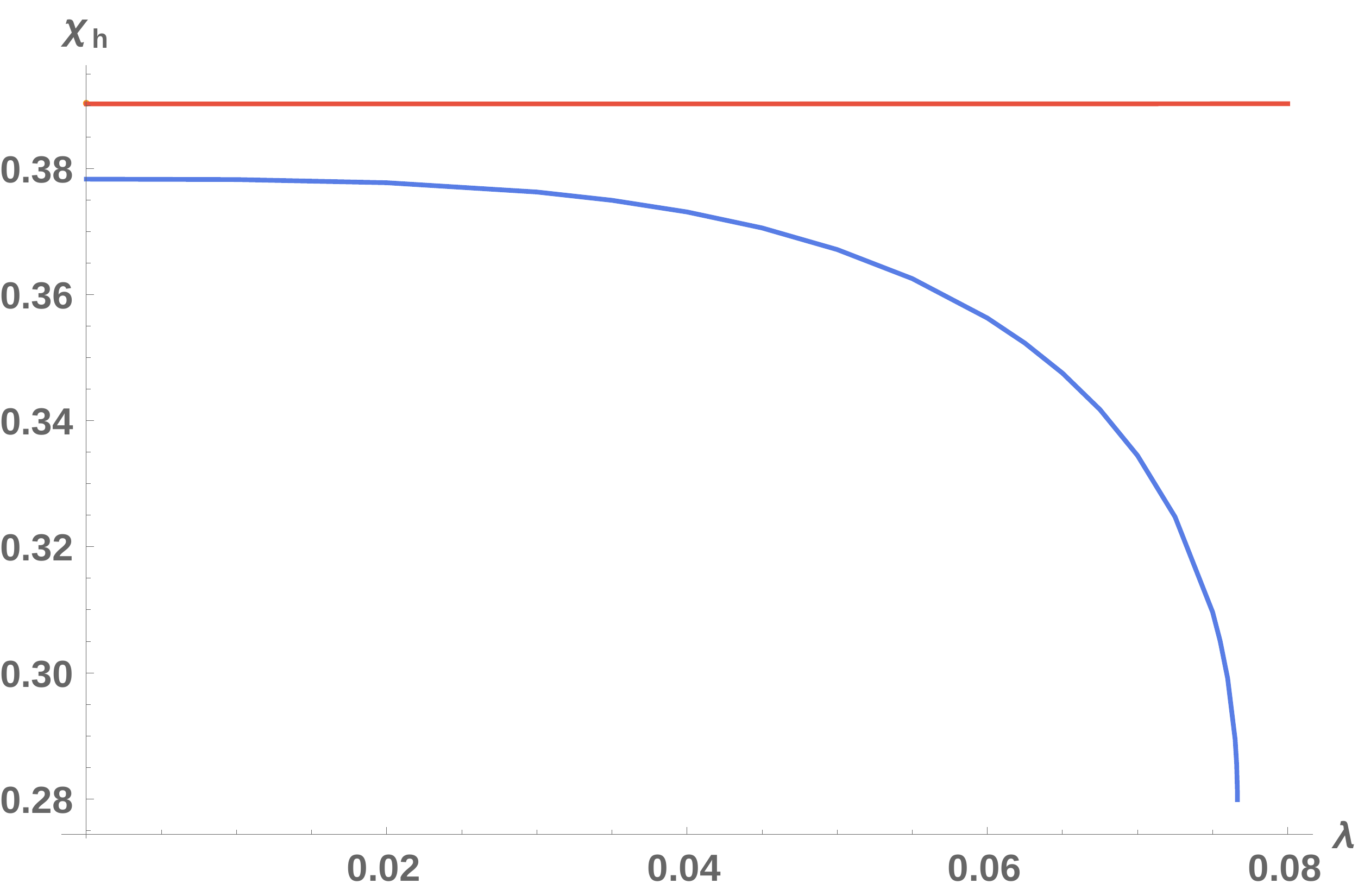}
        \caption{\centering{5 masses.}}
    \end{subfigure}
    \vspace{8mm}
    \begin{subfigure}[t]{0.49\textwidth}
        \includegraphics[width=\textwidth]{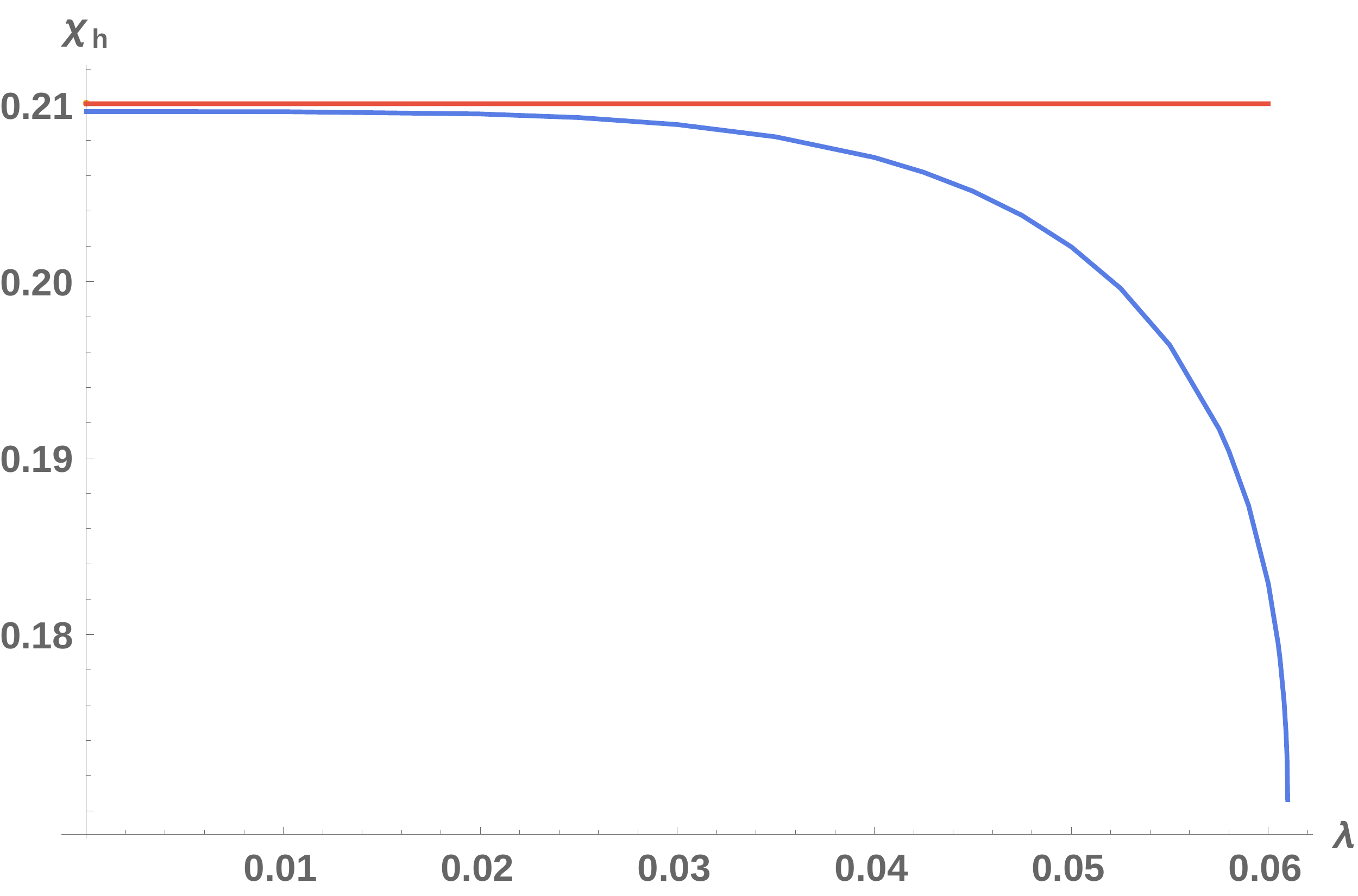}
           \caption{\centering{8 masses.}}
    \end{subfigure}
    \begin{subfigure}[t]{0.49\textwidth}
        \includegraphics[width=\textwidth]{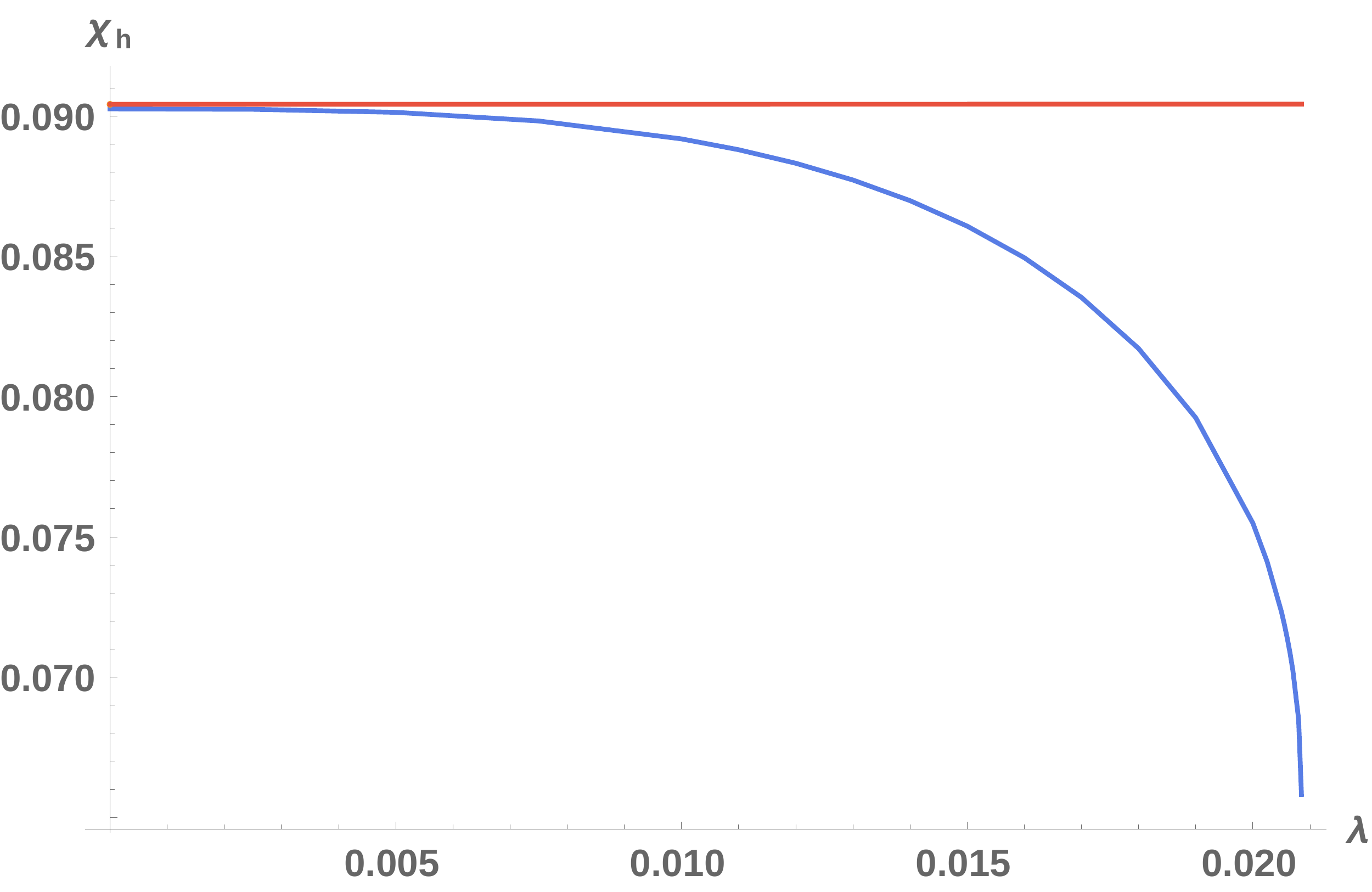}
           \caption{\centering{16 masses.}}
    \end{subfigure}
    \vspace{8mm}
    \begin{subfigure}[t]{0.49\textwidth}
        \includegraphics[width=\textwidth]{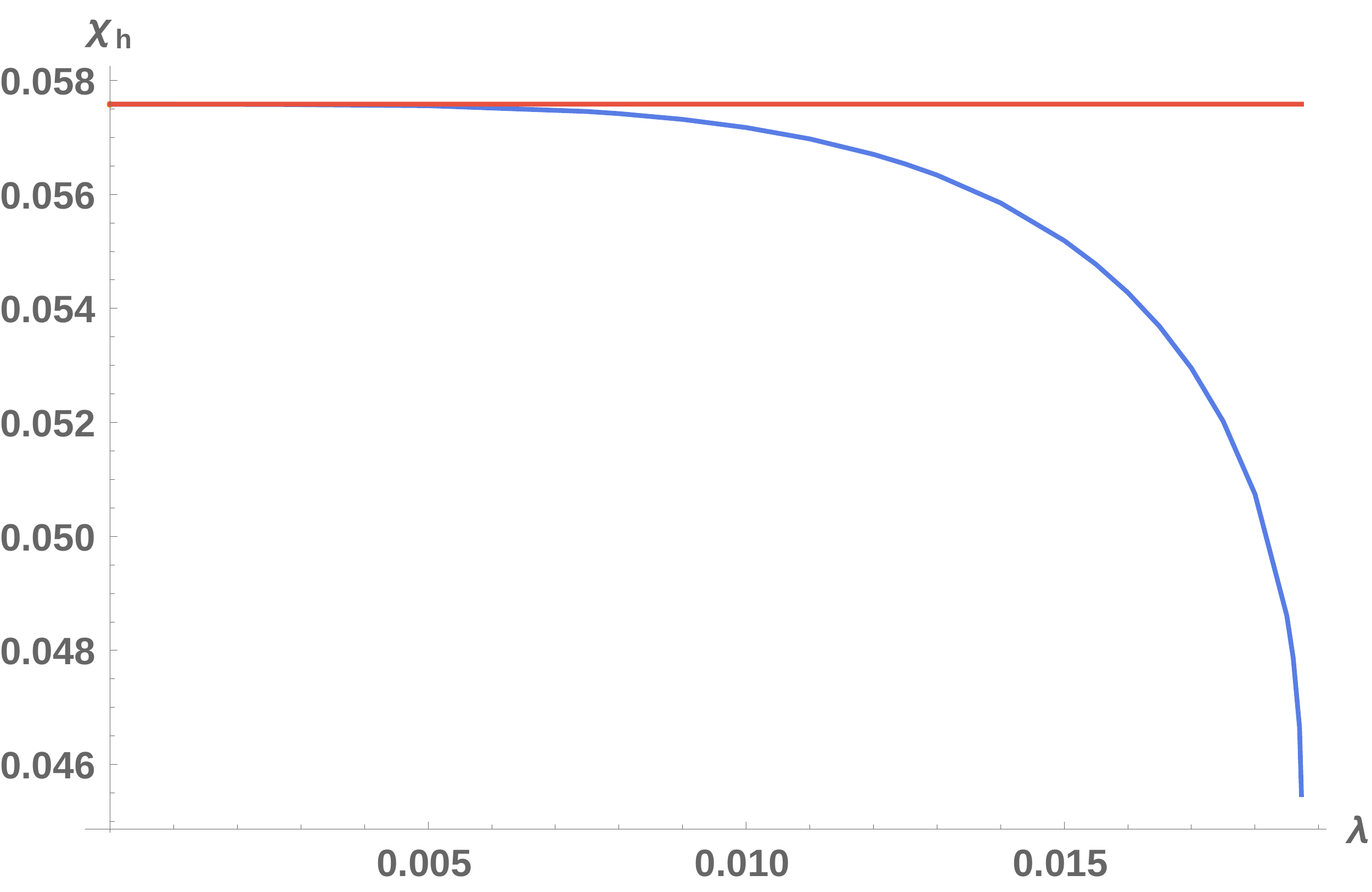}
           \caption{\centering{24 masses.}}
    \end{subfigure}
        \begin{subfigure}[t]{0.49\textwidth}
        \includegraphics[width=\textwidth]{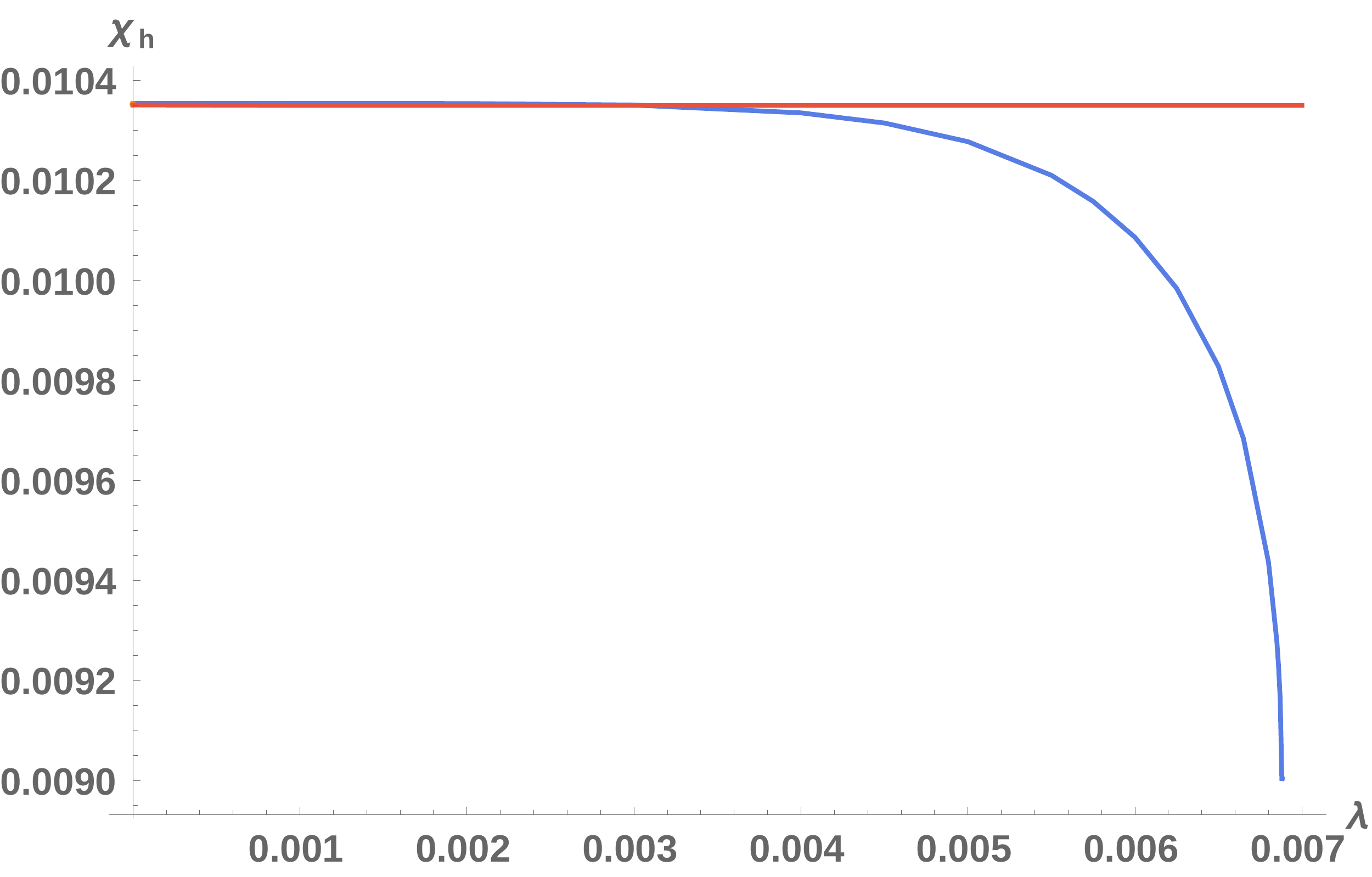}
        \caption{\centering{120 masses.}}
    \end{subfigure}
    \vspace{8mm}
    \begin{subfigure}[t]{0.49\textwidth}
        \includegraphics[width=\textwidth]{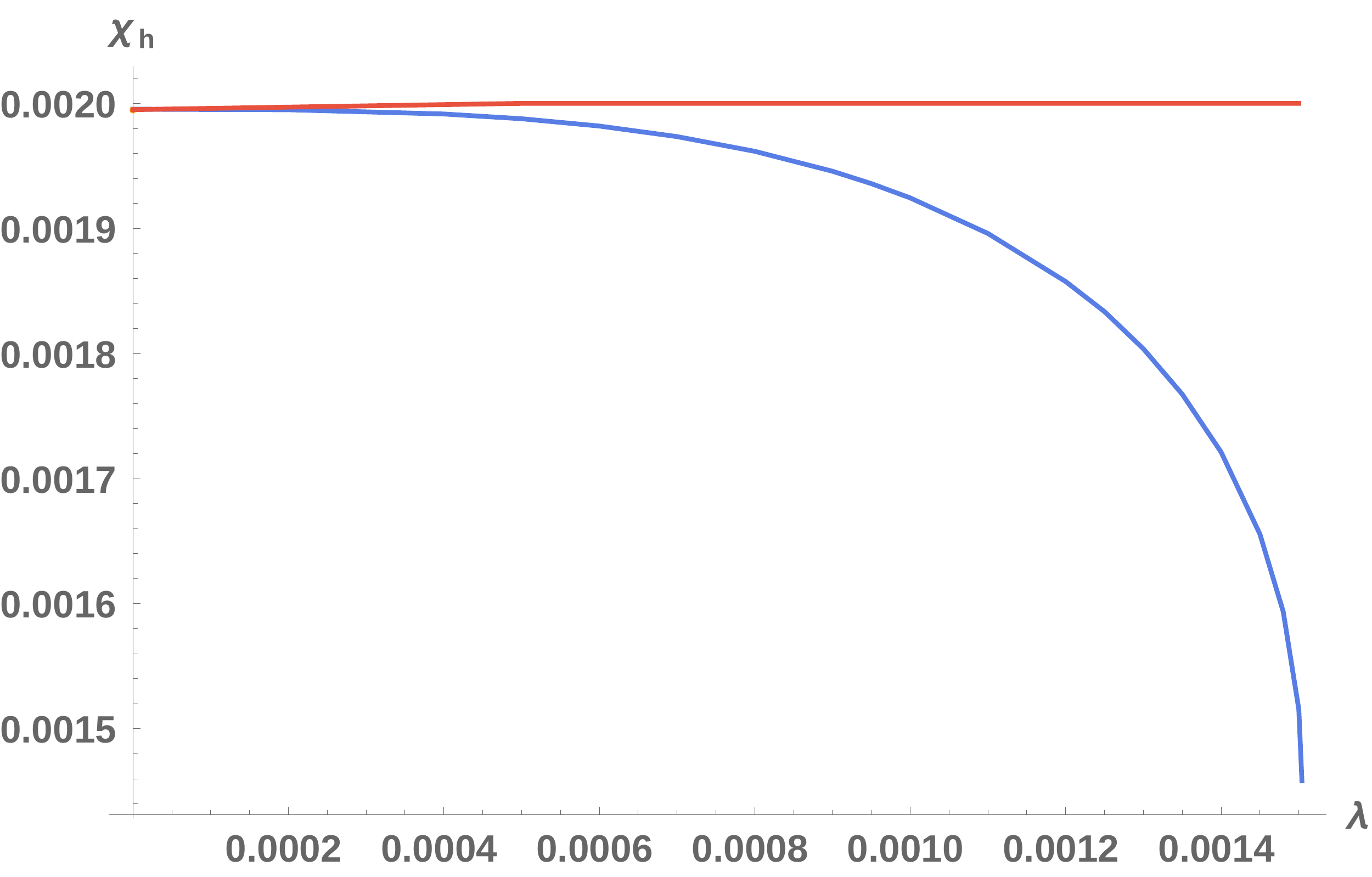}
           \caption{\centering{600 masses.}}
    \end{subfigure}
     \caption{The value of $\rchi$ for the shared horizon at different values of $\lambda$, for the clustered masses around the positions of each of the original black holes in the 6 regular lattices. The horizon position is calculated using both the area method (red lines) and the Weyl tensor method (blue lines).}
    \label{fig:horizons}
\end{figure}

The value of $\lambda_{\rm crit}$ can be calculated for each of the models we are considering, as in each case there is a shared horizon when the masses are tightly clustered on the conformal 3-sphere, which disappears when they become sufficiently separated. These values are displayed in Table \ref{tab:crit} for each of the original lattice structures, and are shown graphically in Fig. \ref{fig:fraction}. It is clear that as the number of masses increases, the value of $\lambda_{\rm crit}$ decreases, reaching values of $\lambda \lesssim 0.01$ when $n \gtrsim 100$. This is to be expected as the relative contribution from each individual black hole to the geometry in the cosmological region decreases for increasing $n$, so the black holes only need to move a small distance apart before the collective horizon disappears when $n$ is large. Figure \ref{fig:horizons} shows the result of repeating this procedure for all of the remaining lattice structures, for the shared horizons. As the number of masses increases, the radius of these horizons decreases due to the decreasing relative contribution to the geometry from each mass for the larger $n$ model. 

The existence of $\lambda_{\rm crit}$ is of interest for understanding the mathematical structure of the horizons in this space-time, but it also has a direct physical relevance for the use of these geometries as initial data for cosmology. This is because observers in the cosmological region will find themselves in universes with different numbers of black holes, depending on the relative magnitude of $\lambda$ and $\lambda_{\rm crit}$. For example, in the 5-mass model (which generalises to a 20-mass model if $\lambda >0$) the observer in the cosmological region will only be able to infer the existence of 5 black holes if $\lambda < \lambda_{\rm crit}$. This is because the additional masses are hidden behind the shared horizons, which is a region of space that should not be expected to be accessible to this observer. On the other hand, if $\lambda > \lambda_{\rm crit}$ then the shared horizons are not present. In this case all 20 masses are distinct, and the cosmological observer will find themselves in a universe that contains all 20 black holes. These are qualitatively different scenarios, which is a point that will become important in Section \ref{sec:scales}, when we compare these geometries to time-symmetric slices through dust-filled FLRW models.

\begin{table}[b!]
\label{tab:crit}
\centering
\begin{tabular}{ |c|c| }
 \hline
  & \\[-5pt]
  $N^{\underline{o}}$ of masses & Critical value at which \\
in original lattice & shared horizons vanish, $\lambda_{\rm crit}$  \\[5pt]
 \hline
  & \\[-5pt]
5  & 0.077 \\
8  & 0.061 \\
16  & 0.021 \\
24 & 0.019\\
120 & 0.007\\
600 & 0.002\\[5pt]
 \hline
\end{tabular}
\caption{\label{tab:crit} Numerical values for $\lambda_{\rm crit}$, to three decimal places.}
\end{table}

\section{Mass Parameters and Interaction Energies}
\label{sec:mass}

The problem of how to define mass in cosmology is a difficult one, as cosmological models (by construction) do not have asymptotically flat regions from which to view the gravitational fields of masses or clusters of masses. Nevertheless, we do need to be able to assign some concept of mass to structures in the Universe if we are to compare our models to Friedmann cosmologies. Complications arise in making these definitions because gravitational interaction energies should themselves be expected to gravitate in Einstein's theory \cite{brillandlind}, and because it is not always entirely clear how to calculate these quantities in cosmology. Indeed, it has recently been suggested that complex hierarchies of structure could produce large deviations from the expectations of Friedmann cosmology, essentially because the cumulative effect of gravitational interaction energies over many different spatial scales could contribute substantially to the total amount of mass in the Universe\footnote{Taking this argument even further, one could make a strong case for stating that almost all of the mass of visible matter in the Universe is due to interaction energies, given that the QCD binding energy accounts for $\sim 99\%$ of all mass in baryons.} \cite{korz2}. We will consider this possibility in the context of our models of clustered black holes. The discussion of mass parameters and interaction energies that follows will proceed as in Refs. \cite{tim3} and \cite{matterincosmo}.

\subsection{Measures of mass in cosmology}

In the work that follows we will consider the following three definitions of mass: bare mass, proper mass, and cluster mass. These are defined as follows:

\paragraph{Bare mass:} The first mass parameter we wish to consider, which has already been introduced in Eq. (\ref{psi0}), is the bare mass of each of our mass points, $\tilde{m}_i$. This particular parameter is introduced by simple analogy with the terms that appear in a time-symmetric slice through the Schwarzschild solution, but does not necessarily correspond in any direct way to the mass that an observer near a black hole would infer for that object. In the case of time-symmetric initial data describing a cluster of black holes in an asymptotically flat space, it is known that the sum of the bare mass parameters corresponds to the sum total of individual proper masses of all the black holes, as well as the sum total of all gravitational interaction energies between each pair of black holes \cite{brillandlind}. The interpretation in the present case is a little more complicated, but we still wish to use these parameters as our first measure of mass. Their meaning will be investigated further below.

\paragraph{Proper mass:} The proper mass of each black hole is defined by rotating our coordinate system so that the black hole we are investigating is centered at $\rchi=0$, and by looking at the behaviour of the geometry in the limit $\rchi \rightarrow 0$. By comparing the leading-order terms in the metric to the Schwarzschild geometry it is then possible to read off a value for the mass, which for a universe containing $N$ black holes is given by
\begin{equation}\label{eq:pmassN}
    m_i = \sum_{j \neq i}^{N} \frac{\sqrt{\tilde{m}_i \tilde{m}_j}}{\sin\left(\frac{\rchi_{ij}}{2}\right)} \, ,
\end{equation}
where $i$ is the label that gives the mass point under consideration, $j$ labels all other masses, and $\rchi_{ij}$ is the coordinate distance between points $i$ and $j$ (after rotating so that $i$ appears at $\rchi=0$). We will refer to $m_i$ as the proper mass of $i$th black hole, which technically corresponds to the mass that an observer in the asymptotically flat region on the far-side of the Einstein-Rosen bridge would infer by looking at how the gravitational field drops off at infinity. This gives a well-defined mass to each of the black holes, but again does not necessarily correspond to the mass that an observer in the cosmological region of the space would infer. It can be seen that the proper mass of a black hole is specified by the positions of all of the other $N-1$ masses in the universe.

\paragraph{Cluster mass:} The third mass parameter we wish to introduce is what we will refer to as the {cluster mass}, $\mathfrak{m}_i$, which will only be defined for black holes that are within a cluster. The idea here is to consider the proper mass that such a black hole would have if we include only contributions to Eq. (\ref{eq:pmassN}) that come from masses outside of the cluster, so that
\begin{equation}\label{eq:pmassn}
    \mathfrak{m}_i = \sum_{j = 1}^{N- C_n} \frac{\sqrt{\tilde{m}_i \tilde{m}_j}}{\sin\left(\frac{\rchi_{ij}}{2}\right)} \, ,
\end{equation}
where again $i$ labels the black hole under consideration, which we arrange to be at $\rchi_i = 0$, but this time $j$ labels the masses outside of the cluster. The $N$ is again the total number of black holes in the universe, and  $C_n$ here corresponds to the number of masses in each cluster (which, from our construction, is equal to the number of vertices in the primitive lattice cell of the original $n$-mass lattice). We therefore have that $N=n \,  C_n$, and the summation in Eq. (\ref{eq:pmassn}) can be seen to be over the $N-C_n$ black holes outside the cluster under consideration. 

\paragraph{} These three definitions of mass will be used below, to give measures of the gravitational interaction energies both within and between clusters of masses. We expect this to clarify the reasons for using these particular definitions. We will also use them in Section \ref{sec:scales}, to enable us to compare our models to Friedmann cosmologies. For further discussion of masses and interaction energies between black holes in cosmological models the reader is referred to Refs. \cite{tim3} and \cite{matterincosmo}.

\subsection{Interactions within and between clusters}
\label{sec:int}

Our motivation for considering gravitational interaction energies is that, in Einstein's theory, they themselves are known to act as sources of mass. For example, for two black holes of mass $m_1$ and $m_2$, separated by distance $r$ in an asymptotically flat space, the total gravitational mass of the system when viewed from infinity is given, to leading-order, by \cite{brillandlind}
\begin{equation}
\label{eq:brill}
M_{\rm tot} = m_1 + m_2 -\frac{m_1 m_2}{r} \, .
\end{equation}
In general, for many masses, the expression for the total mass of the whole system becomes
\begin{equation} \label{MMM}
M_{\rm tot} = M_{\rm obj} +  M_{\rm int} \, ,
\end{equation}
where $M_{\rm obj}$ is the sum total mass of all of the individual objects in the system, and $M_{\rm int}$ is the sum total of all of the pairwise interactions between objects in the system. Applying this formula to the black holes within a cluster, and the clusters within the cosmology, will allow us to determine values for the relevant gravitational interaction energies both within and between clusters of black holes.

\paragraph{Interaction energies within clusters:} If we start by considering the individual black holes within a cluster, then we can derive an expression for the interaction energies between all objects within that cluster, $M_{\rm int}^{\rm wcl}$. This was calculated in Ref. \cite{tim3}, and is given by
\begin{equation}
M_{\rm int}^{\rm wcl} \simeq - \sum_i^{C_n} \sum_{j \neq i}^{C_n-1} \frac{\sqrt{\tilde{m}_i\tilde{m}_j}}{\sin({\frac{\chi_{ij}}{2}})} \, ,
\end{equation}
where the sum over $i$ is over all the $C_n$ masses in the cluster, and the sum over $j$ is over the remaining $C_n - 1$ masses in the cluster (for every $i$th black hole). Using this expression, we can write the sum of intra-cluster interaction energies, $M_{\rm int}^{\rm wcl}$, in terms of the mass parameters in the previous section as
\begin{align}
     M_{\rm int}^{\rm wcl} =M_{\rm tot}^{\rm cl}-M^{\rm bh}_{\rm obj}
     \label{eq:mwcl:2}
     & \simeq \sum_i^{C_n} \mathfrak{m}_i - \sum_i^{C_n} m_i \, ,
     \end{align}
where in this case $M^{\rm bh}_{\rm obj}$ is the sum total of proper masses of the individual black holes in the cluster, and where $M_{\rm tot}^{\rm cl}$ is the total mass of the cluster (which can now be seen to be given by the sum of $\mathfrak{m}_i$ over all black holes within the cluster).

\paragraph{Separation interaction energies:} Let us now consider the interaction energies between black holes separated by cosmological distances, and the total mass of the universe as a whole. These quantities are difficult to define, but a natural definition for the total energy in the cosmological model is given by \cite{tim}
\begin{equation}
M_{\rm tot}^{\rm cos} = \sum_i^{N} \tilde{m}_i \, .
\end{equation}
This quantity is defined in analogy to the standard expression for $N$ black holes in an asymptotically flat space \cite{brillandlind}, and is closely related to the expression that one would obtain when considering the total mass of all black holes and interaction energies in one of the asymptotically flat flange regions on the far side of an Einstein-Rosen bridge \cite{matterincosmo}. Using it to evaluate the inter-cluster gravitational interaction energies, we find
\begin{equation} \label{Mintsep}
M_{\rm int}^{\rm sep} = M_{\rm tot}^{\rm cos} - n \, M_{\rm obj}^{\rm bh}   = \sum_i^N \tilde{m}_i - \sum_i^N m_i \, ,
\end{equation}
where the term containing $M_{\rm obj}^{\rm bh}$ again indicates the total proper mass of all black holes, but this time includes all $N$ black holes in the universe. This expression has an appealing similarity to the one given in Eq. (\ref{eq:mwcl:2}), and is intended to give a measure of the total of all gravitational interaction energies between all black holes.

\paragraph{Interaction energies between clusters:} Using the results above, we can deduce an expression for the interaction energies between clusters that are separated by cosmological scales. Using Eq. (\ref{MMM}) we find this quantity to be equal to
\begin{equation}
M_{\rm int}^{\rm bcl} = M_{\rm tot}^{\rm cos} - n\, M_{\rm tot}^{\rm cl} \, ,
\end{equation}
where $M_{\rm tot}^{\rm cos}$ is again the total energy in the entire model, and $M_{\rm tot}^{\rm cl}$ here is the total energy within each cluster (i.e. proper masses of the black holes plus intra-cluster interaction energies). From our previous considerations, this can be written as
\begin{equation} \label{Mintbcl}
M_{\rm int}^{\rm bcl} \simeq \sum_i^N \tilde{m}_i - n \sum_i^{C_n} \mathfrak{m}_i = \sum_i^N \tilde{m}_i - \sum_i^{N} \mathfrak{m}_i \, ,
\end{equation}
where $n$ is the number of clusters (equal to the number of masses in the original lattice), and where in the last equality we have used the fact that all clusters are identical. This expression is intended to describe the sum total of all gravitational interaction energies between all clusters of black holes. Again, it has a pleasing similarity with (\ref{eq:mwcl:2}) and (\ref{Mintsep}).

\begin{figure}[tbp]
    \centering
    \begin{subfigure}[b]{0.495\textwidth}
        \includegraphics[width=\textwidth]{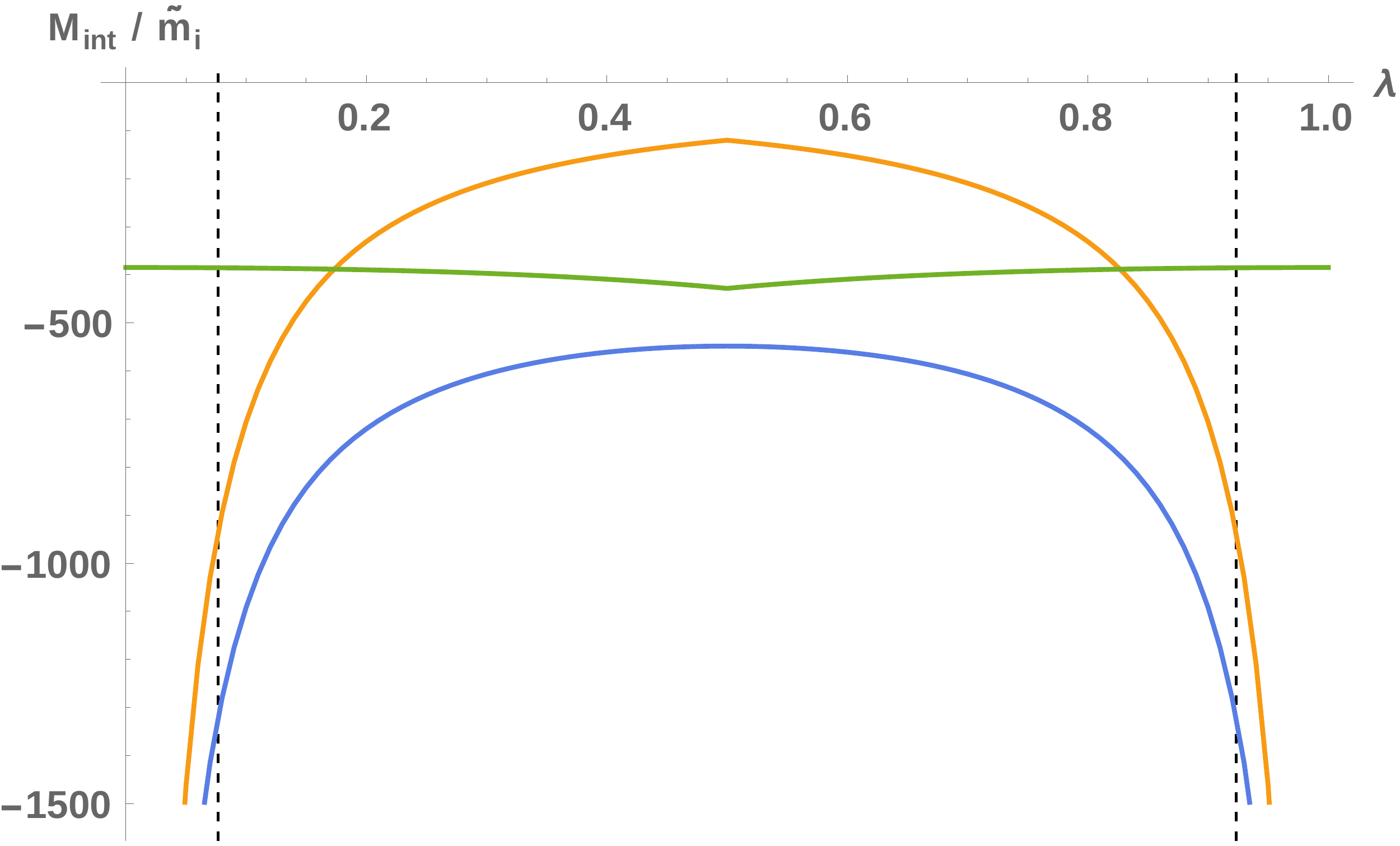}
        \caption{\centering{5 $\to$ 5 masses.}}
    \end{subfigure}
    \vspace{2mm}
    \begin{subfigure}[b]{0.495\textwidth}
        \includegraphics[width=\textwidth]{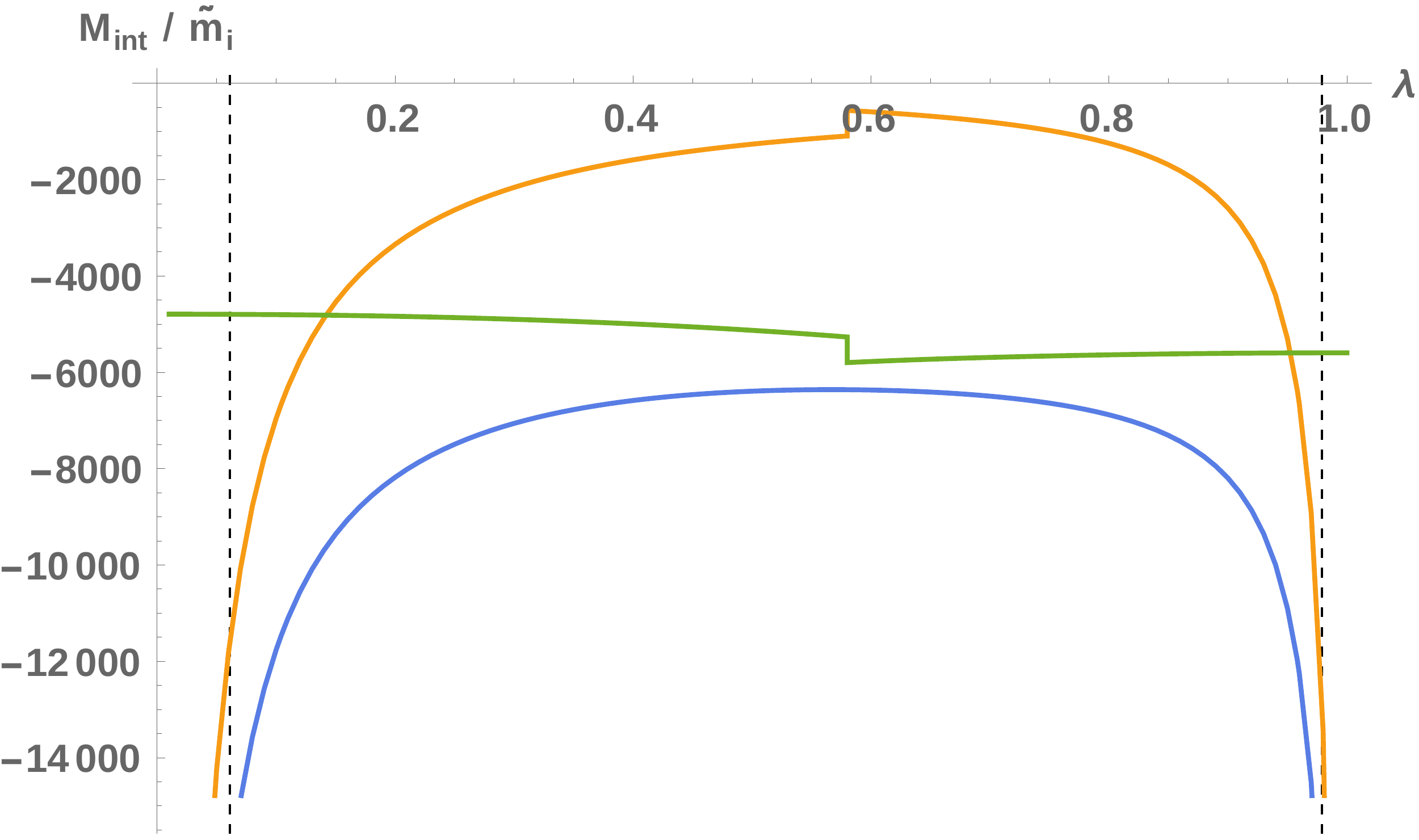}
           \caption{\label{fig:int:2}\centering{8 $\to$ 16 masses.}}
    \end{subfigure}
    \begin{subfigure}[b]{0.495\textwidth}
        \includegraphics[width=\textwidth]{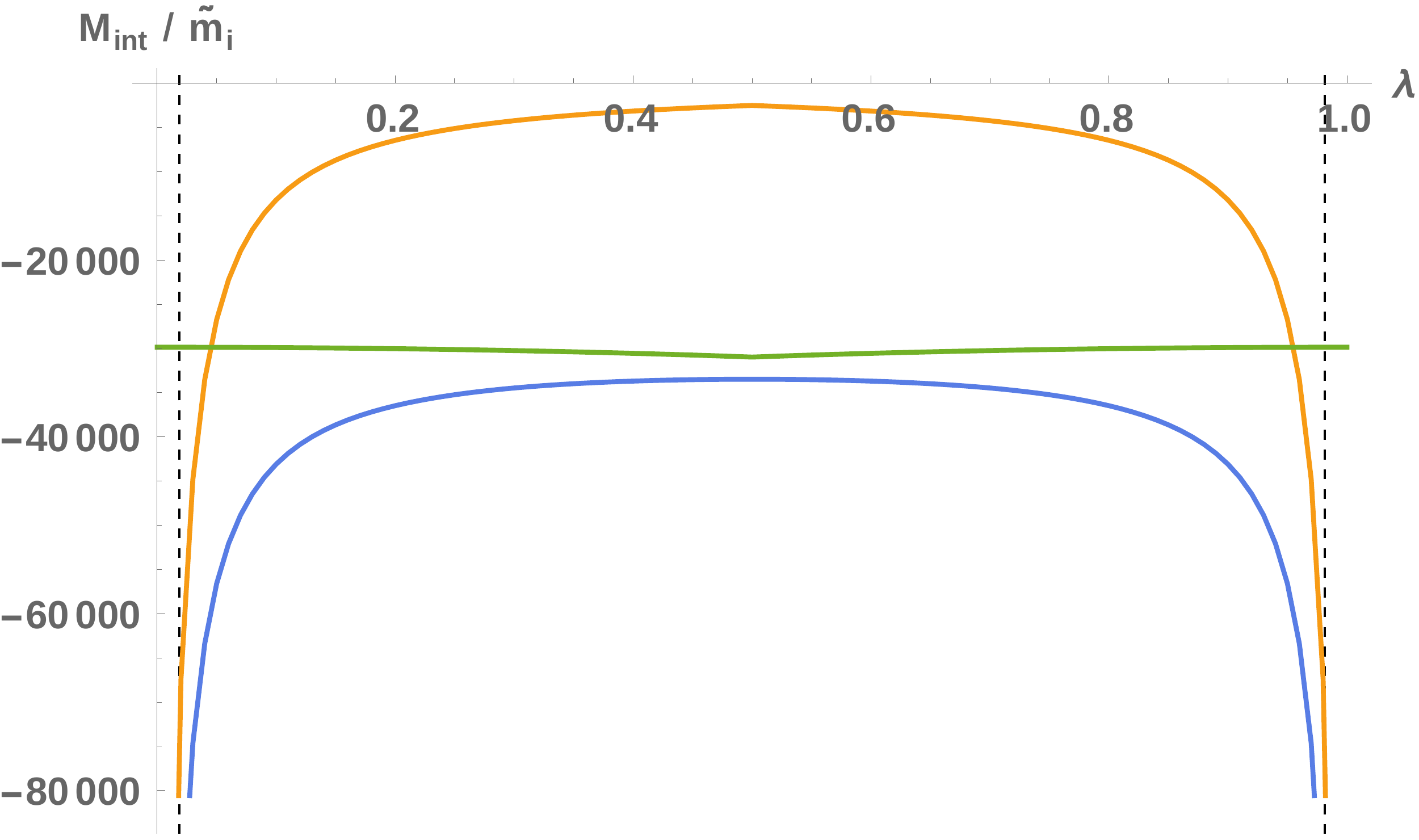}
           \caption{\centering{24 $\to$ 24 masses.}}
    \end{subfigure}
    \vspace{2mm}
    \begin{subfigure}[b]{0.495\textwidth}
        \includegraphics[width=\textwidth]{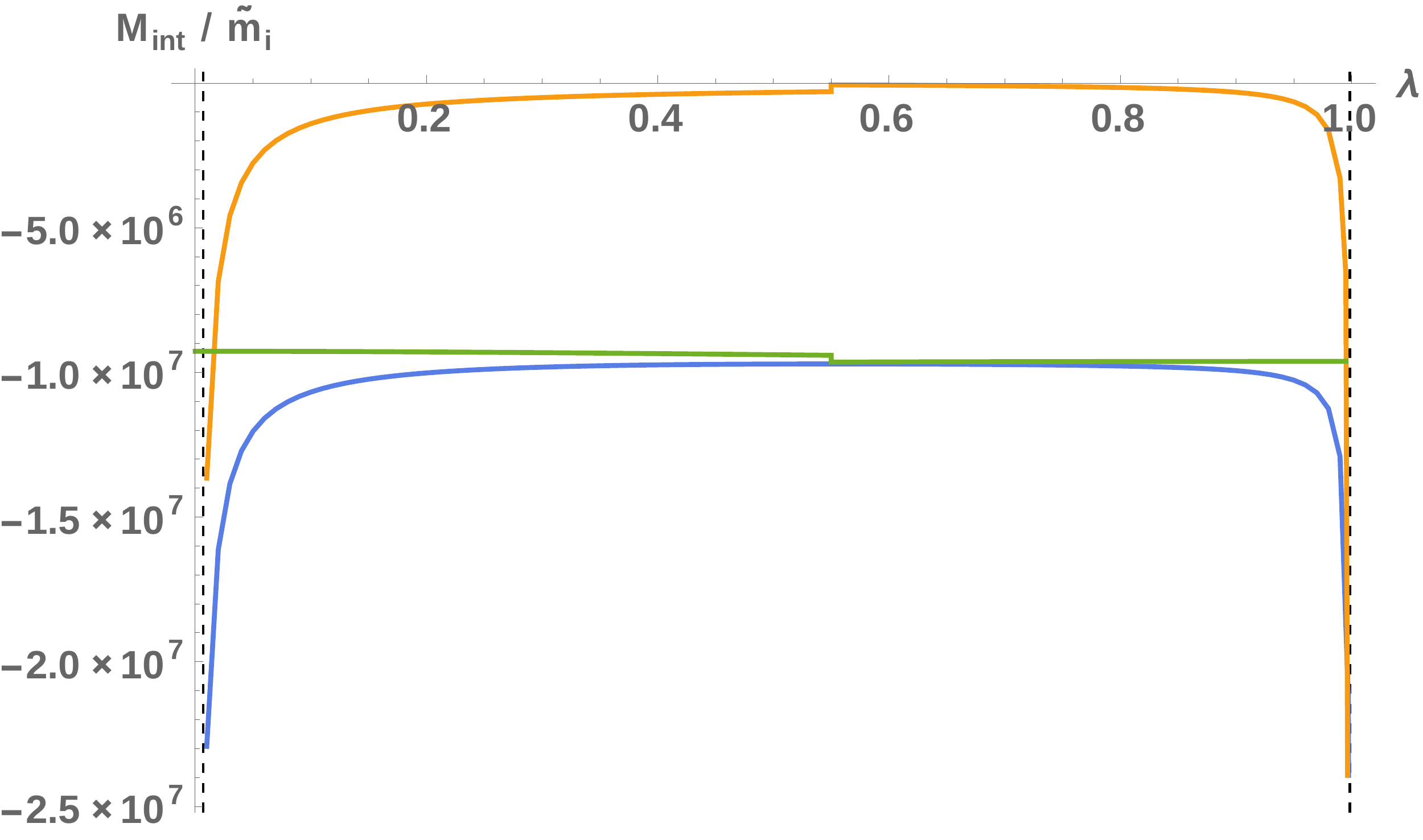}
           \caption{\label{fig:int:4}\centering{120 $\to$ 600 masses.}}
    \end{subfigure}
     \caption{All three interaction energies for each of the configurations under consideration, as a function of $\lambda$, and in units of mass parameter $\tilde{m}_i$. The blue lines correspond to the total separation interaction energy, $M_{\rm int}^{\rm sep}$, the orange lines give the interaction energies within $n$ clusters, $n \, M_{\rm int}^{\rm wcl}$, and the green lines give the interaction energy between clusters, $M_{\rm int}^{\rm bcl}$.}
    \label{fig:int}
\end{figure}

\paragraph{} Let us now consider these different types of gravitational interaction energy, within the models discussed in Section \ref{sec:structure}. For simplicity, and in each model, all of the mass parameters of all of the black holes will be taken to be identical. When the black holes are clustered, around one of the cell centres, there will of course be interactions between the $C_n$ black holes within each cluster, as well as interactions between the clusters. These two interaction energies are depicted as the orange and green lines in Fig. \ref{fig:int}, respectively. Their sum, given by the blue line, is the total of all interaction energies between all black holes in the model. It is clear from the figure that the intra-cluster interaction energies diverge as $\lambda \rightarrow 0$, in the limit that the black holes get closer together. The total interaction energy between all separated black holes also diverges, but the inter-cluster interactions defined in Eq. (\ref{Mintbcl}) can be seen to remain finite and close to constant. This is exactly what we would expect from these quantities, and gives some {\it a posteriori} justification for their introduction.

The four plots in Figure \ref{fig:int} are sufficient to give all the information about all of the interaction energies in all of our clustered-mass models. This is, for example, because the models that contain masses clustered around the positions of the cell centres of the original 8-mass lattice are given by the $\lambda < 1/2$ portion of plot (b), while models with masses clustered around the positions of the cell centres of the original 16-mass lattice are given by the $\lambda >1/2 $ portion of the same plot. This is due to these two lattices being dual to each other, with the 8-mass model recovered in the limit $\lambda \rightarrow 0$ and the 16-mass model recovered in the limit $\lambda \rightarrow 1$. Similarly, plot (d) is sufficient to describe black holes clustered around the positions of the cell centres in the original 120 and 600-mass models. In the case of plots (a) and (c) only clustered masses around one type of lattice cell centres are considered (the 5 and 24-mass models, respectively). This is because the 5 and 24-cell lattices are both self-dual.

The critical values of $\lambda$ are denoted in each of these plots by the vertical dashed lines. Recall that these values of $\lambda$ are physically very important, as they denote when the clustered masses become distinguishable as individual black holes to observers in the cosmological region. The reader may note that the halfway point between original setups occurs at $\lambda = 0.5$ only for the models that are self-dual (i.e. for the $5$ and $24$-mass models). Due to the asymmetry between transitioning from an $8$-mass model to a $16$-mass model, however, the halfway point (when individual black holes are maximally separated) is instead at $\lambda \simeq 0.576$. For the transition from $120$ to $600$ masses this point occurs at $\lambda \simeq 0.553$. These are the values of $\lambda$ at which the glitches occur in the green and orange curves in plots (b) and (d) -  they are due to changing whether we consider the black holes to be clustered around the centres of the lattice cells or the dual lattice cells. Of course, for values of $\lambda \sim 1/2$ these masses should not really be considered as being clustered around either point, so this discrepancy has no real practical significance.

The reason for introducing and studying these different interaction energies will become clear in the next section, when we compare our clustered-mass models to Friedmann solutions and to each other. In order to do this quantitatively, we need to know which (if any) of the interaction energies should be included in the total mass content of our black hole universes.

\section{Cosmological Consequences}
\label{sec:scales}

We now wish to evaluate the consequences of the clustering of black holes on the scale of the cosmological region. To perform this study requires having either reference models or having some way of comparing one lattice universe to another. In Section \ref{compFLRW} we will compare our lattice models to FLRW models that are, in some sense, similar to our black hole universes. This will require addressing a number of subtleties, which we discuss in detail as we present our results. In Section \ref{complattice} we then present a comparison between lattice models that have different levels of structuration (i.e. different values of $\lambda$), but the same total energy content. 

\subsection{Comparison with FLRW}
\label{compFLRW}

In order to compare our models to FLRW cosmologies we need to know which FLRW model we should choose. This is a more subtle problem than it may initially appear. The objective of such a comparison is to allow us to consider the effect that the existence of structure has on the large-scale cosmology. This suggests choosing an FLRW model that has a matter content composed of pressureless dust, with the knowledge that this is the type of matter field routinely used in studies of Friedmann cosmology to model the gravitational effects due to astrophysical objects such as galaxies and clusters of galaxies. The time-symmetry of our initial hypersurface also suggests choosing an FLRW model that contains a similarly time-symmetric 3-dimensional subspace. This means models where the spatial curvature constant is given by $k=+1$. The final remaining question is how much mass should exist in this positively-curved dust-dominated Friedmann model, which is much more difficult to answer uniquely. We will consider different prescriptions for this in what follows, in order to open the question to further scrutiny.

One way to directly compare between the two types of universe is to look at a measure of global scale, which we will choose to be the scale factor in the cosmological regions of each space. For an FLRW at maximum of expansion, the scale factor can be read off from the line-element \cite{tim}
\begin{equation}
\label{eq:lineelement}
ds^2 = \frac{16 M^2}{9 \pi^2}\left( d\rchi^2 + \sin^2\rchi d\Omega^2 \right),
\end{equation}
as
\begin{equation}
a^F(t_0) = \frac{4 M}{3 \pi} \, ,
\end{equation}
where $M= \rho(t_0) V$ is just the total mass of dust with energy density $\rho(t_0)$ in an FLRW universe of spatial volume $V$. There is no ambiguity in choosing this scale factor, as the FLRW models are by construction identical at every point in space and in every direction. Choosing an appropriate scale factor in our black hole universes is a more ambiguous process.

If we consider the form of line-element given in Eq. \eqref{eq:conformal}, then the scale factor of the lattice models is simply
\begin{equation}
a^L(t_0) = \psi^2(\rchi,\theta,\phi) \, ,
\end{equation}
where $\psi$ is given by Eq. (\ref{psi0}). The difficulty of trying to identify a value for this quantity is that $\psi$ is a function of position on the conformal 3-sphere. One must therefore choose one particular position, to get a unique value for $a^L(t_0)$. We should obviously choose a position in the cosmological region of the space, if we wish to get a measure of the scale factor of the cosmology. More specifically, and in order to make this choice as unambiguous as possible (as well as to minimise the effect of any individual nearby black hole), we choose the point that is furthest away from all masses, where $\psi$ is at its global minimum. The resulting scale factor is then uniquely specified for any value of $\lambda \in [0,1]$.

The reader may note that this is a different definition of scale to that which is often chosen in models that contain regular arrays of black holes (see e.g. \cite{evolution3}). In those cases it is a convenient and well-defined process to calculate the proper length of a lattice cell edge. Such curves are maximally far from all black holes (by construction), and are uniquely picked out by the geometry. In our case this would not be a good choice for a measure of scale, as when $\lambda \rightarrow 1$ the black holes themselves end up at the ends of these curves. This would result in a measure of scale that involved integrating through regions of space that were outside of the cosmological region, and that itself diverges in the limit $\lambda \rightarrow 1$. This is obviously undesirable, which is why have chosen to use the scale factor described above. A similar method was used in the regular lattice case in Ref. \cite{tim}, and the results were found to be comparable to the lattice cell-edge method.

\begin{figure}[t!]
    \centering
  \includegraphics[width=0.75\textwidth]{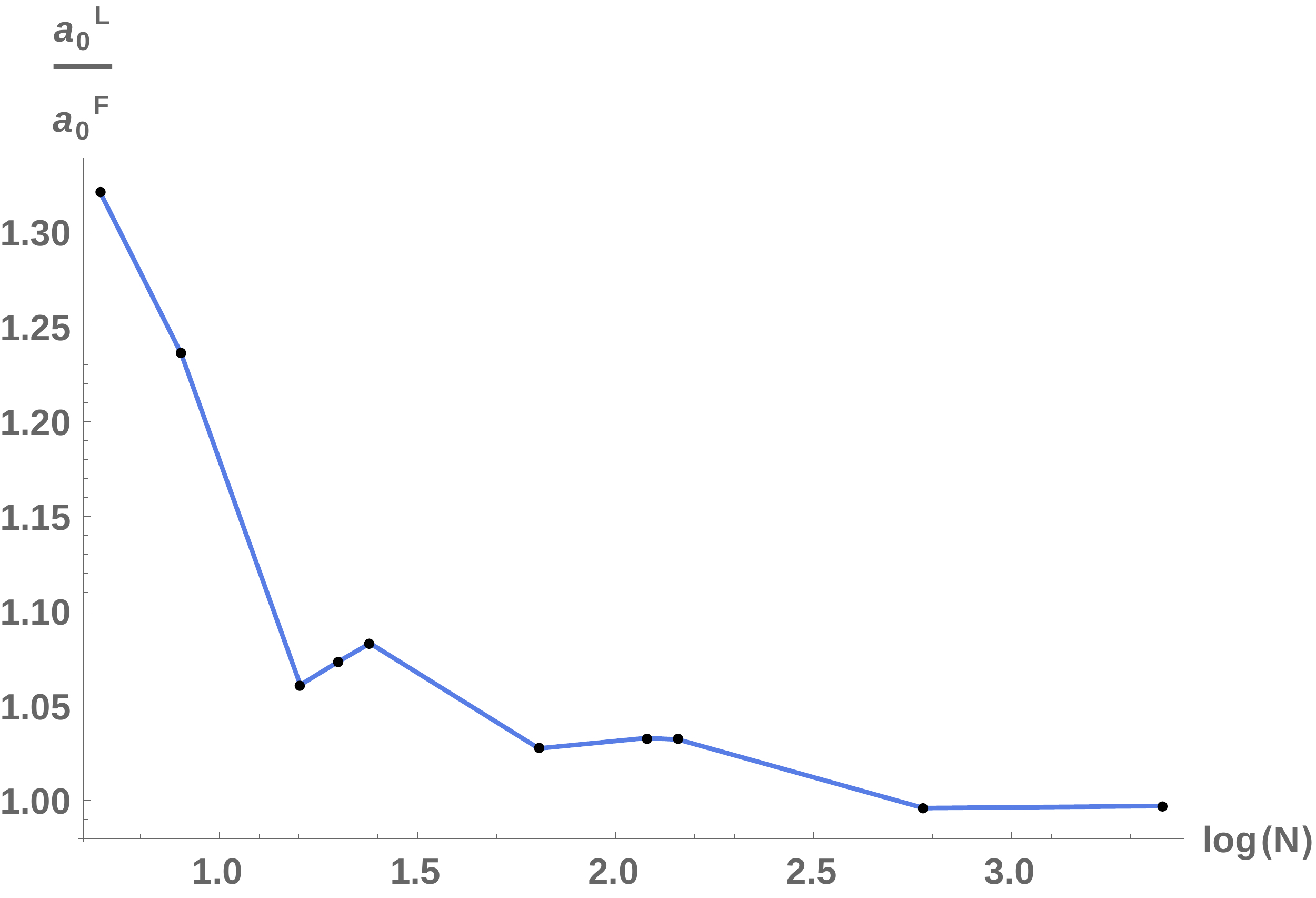}
     \caption{\label{fig:scales} The ratio of scale factors in the lattice universes $a_0^L$ and the dust-dominated positively-curved FLRW universes $a_0^F$, as a function of the number of masses in the universe $N$, where $N = 5, 8, 16, 20, 24, 64, 120, 144, 600$ and $2400$.}
\end{figure}

Let us first consider the ratio of these scale factors at values of $\lambda$ that correspond to the points when the masses are all maximally far apart. This obviously includes the original models with $\lambda =0$ and $1$, but now also includes models where the black holes are at `halfway' positions, with $\lambda \sim 0.5$. This will allow us to extend our existing data set of 6 perfect lattices by another 4 configurations that are almost regular. These new models will have 20, 64, 144 and 2400 black holes that are well spaced, but not perfectly regular. The results are plotted in Figure \ref{fig:scales}, and the values of the 10 data points given in Table \ref{tablereg}. In this case we have compared the scale factor from our lattice universes to the scale factor in dust-dominated positively-curved FLRW universes with the same total proper mass, as defined in Eq. (\ref{eq:pmassN}). The general trend is clear, if not entirely uniform; the effect of increasing the number of masses is to reduce the scale of the lattice universes, such that for large $N$ they become increasingly similar to FLRW universes that contain the same total proper mass.

\begin{figure}[t!]
    \centering
  \includegraphics[width=0.88\textwidth]{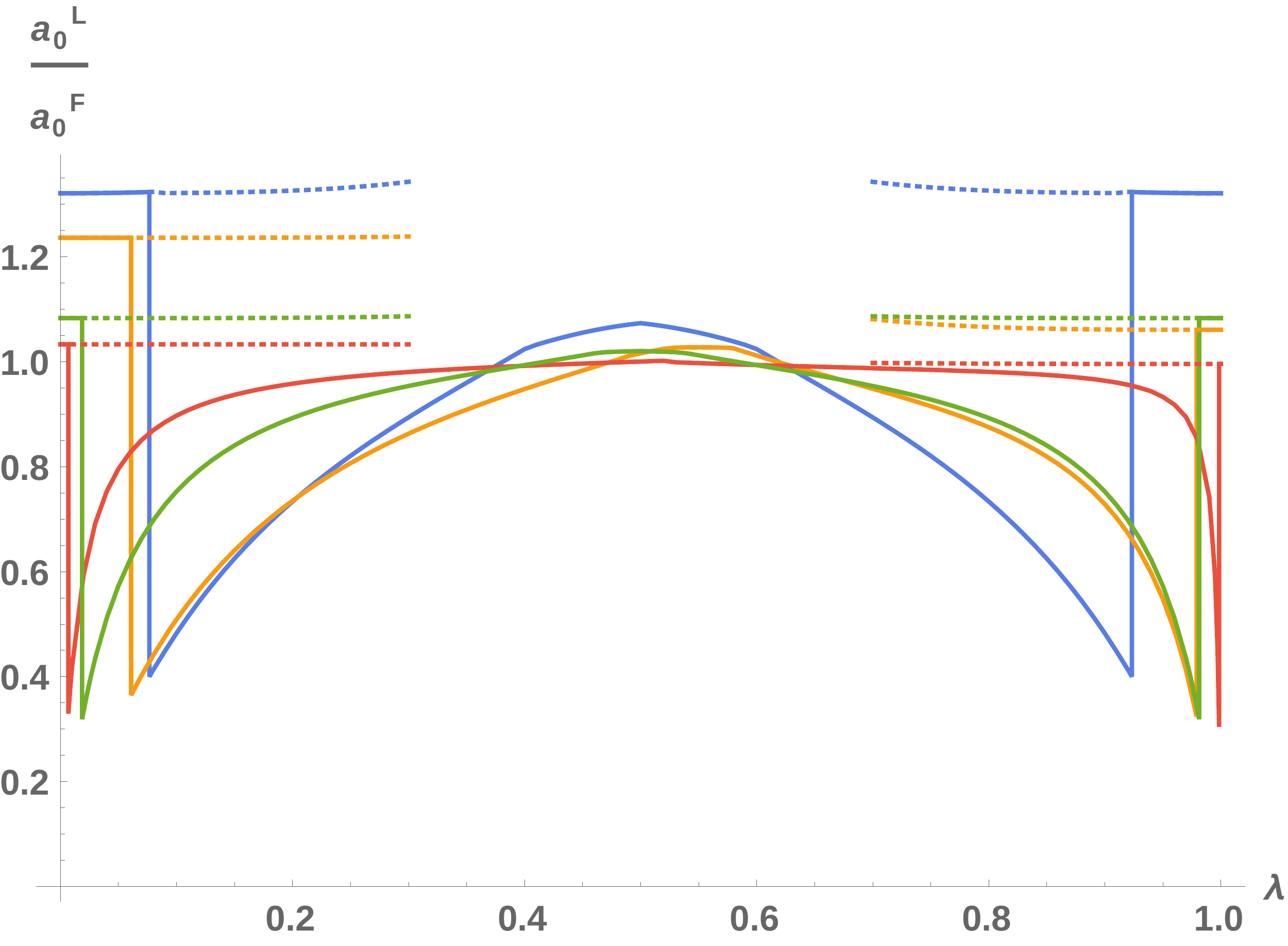}
     \caption{\label{fig:scales2} The ratio of scales between the lattice universes $a_0^L$ and the corresponding dust-dominated positively-curved FLRW universes $a_0^F$, as a function of $\lambda$. The four curves correspond to the models with the following number of masses: $5 \to 5$ (blue), $8 \to 16$ (orange), $24 \to 24$ (green), and $120 \to 600$ (red). The sudden drops appear at the points where $\lambda=\lambda_{\rm crit}$, which have different values for each of the different cases. In the regions where $\lambda$ is beyond $\lambda_{\rm crit}$, the dotted lines correspond to the total mass being calculated using the sum of cluster masses ($\sum_i \mathfrak{m}_i$) and the solid lines correspond to the total mass being calculated using the sum of proper masses of individual black holes ($\sum_i m_i$).}
\end{figure}

Next, we want to consider lattice configurations with all values of $\lambda$, and determine the effect that clustering has on the ratio of scales between lattice and FLRW universes. For this we need to be careful about how we determine the total mass in the lattice models. Recall that if $\lambda$ is within the critical value $\lambda_{\rm crit}$ then the number of black holes in the cosmological region is given by the number of cells in the original lattice model, $n$. If this is the case it is most appropriate to use Eq. \eqref{eq:pmassn} to determine the proper mass of the cluster within the shared horizon, so that $M=\sum_i \mathfrak{m}_i$. This measure includes the interaction energies of the black holes within the shared horizon, which are now indistinguishable from the perspective of a cosmological observer. In this case, we then compare the lattice models to an FLRW universe with the same total mass as the sum of all of these cluster masses. If $\lambda$ is beyond the critical value, however, then there are $N = n \, C_n$ distinct black holes in the cosmological region, where $C_n$ is the number of masses in a cluster. In this case we could choose to calculate the total mass of each cluster of black holes using $\sum_i \mathfrak{m}_i$, or use $\sum_i m_i$ where $m_i$ is the proper mass of each black hole. The former of these methods will include the intra-cluster interaction energies, while the latter will not. We display the results for both methods in Figure \ref{fig:scales2}, for each of our lattice models. 

\begin{table}[b!]
\label{tablereg}
\centering
\begin{tabular}{ |c|c| }
 \hline
  & \\[-5pt]
  $N^{\underline{o}}$ of masses & Ratio of scale factors, for an FLRW universe \\
in black hole universe & with the same total proper mass, $a_0^L/a_0^F$  \\[5pt]
 \hline
  & \\[-5pt]
5  & 1.321 \\
8  & 1.236 \\
16  & 1.061 \\
20  & 1.073 \\
24 & 1.083\\
64 & 1.027 \\
120 & 1.033\\
144 & 1.032\\
600 & 0.996\\
2400 & 0.997\\[5pt]
 \hline
\end{tabular}
\caption{\label{tablereg} Numerical values for the ratio $a_0^L/a_0^F$, to three decimal places.}
\end{table}

The results of using this method show the expected behaviour for the limiting cases when $\lambda \rightarrow 0$ or $1$, as determined in Ref. \cite{tim}. The results for values of $\lambda$ near to these extreme cases also appear well behaved, as expected due to the fact that we are still considering an effective $n$-mass model (the extra masses are behind a shared horizon). This shows that the cluster mass $\mathfrak{m}_i$ is a more appropriate measure of mass to use than the individual proper masses $m_i$ of the points within the shared horizon. If we had chosen to use this latter measure, which naively might have been thought of as appropriate, then the behaviour of $a_0^L/a_0^F$ would have been to diverge away from $1$ as $\lambda \rightarrow 0$ or $1$. This would have been unacceptable, as the geometry in the cosmological region should be smoothly approaching that of the regular lattices in these limiting cases, which means there should be no noticeable effects from having split the original mass into $C_n$ new masses (from the point of view of an observer in the cosmological region). This shows that interaction energies within a shared horizon should be expected to gravitate, and to contribute to the scale of the universe.

If we now consider the values of $\lambda$ beyond the critical values, then we can see from Fig. \ref{fig:scales2} that the comparisons made using proper masses of the black holes (solid lines) can yield quite different results from those made using the cluster mass (dotted lines). If we take the total mass of the black hole universes to be given by the sum total of individual proper masses of all black holes within it, as has frequently been done in previous studies, then the results can be seen to vary strongly with changing $\lambda$. In the most extreme case, which is the 600 mass model, the scale factor of the lattice universe drops by up to 30\% of its FLRW counterpart when the black holes are made to cluster (i.e. when the masses are close together, but still separated enough to be considered individual black holes). However, when the ratio of scale factors are determined using the total cluster masses of the black hole (which include the interaction energies within each cluster), it can be seen that the dependence of the scale factor $a_0^L$ on $\lambda$ is much less pronounced. In Figure \ref{fig:scales2}, the dotted lines that give the ratio of scale factors in this case have been truncated artificially, as at some point it becomes untenable to consider black holes separated by large distances as belonging to a ``cluster''.

These results again suggest that the intra-cluster interaction energies should be expected to contribute to the total mass budget of the universe, even when $\lambda$ is beyond $\lambda_{\rm crit}$. This is because the situations immediately before and after the critical values of $\lambda$ do not produce discontinuous changes in the geometry of the cosmological region when we pass $\lambda_{\rm crit}$, only an instantaneous change in the existence of the shared horizon. From a physical perspective, it therefore seems the case that the appropriate measure of total mass should include the interaction energies that exist within the cluster, even when the shared horizon does not exist. This raises some interesting questions about dynamical structure formation in the real Universe: if interaction energies contribute to the energy budget of the Universe, then which interactions between which pairs of objects should be considered? In the examples considered in this section we have artificially introduced clusters of objects on a single scale. In the real Universe structure exists on a multitude of scales, and pairwise interactions could, in principle, be expected to exist between every pair of objects in the Universe. Should we take each of these interactions into account? And, if not, where should we draw the line?

\subsection{Comparison between lattice models}
\label{complattice}

\begin{figure}
    \centering
    \begin{subfigure}[t]{0.49\textwidth}
        \includegraphics[width=\textwidth]{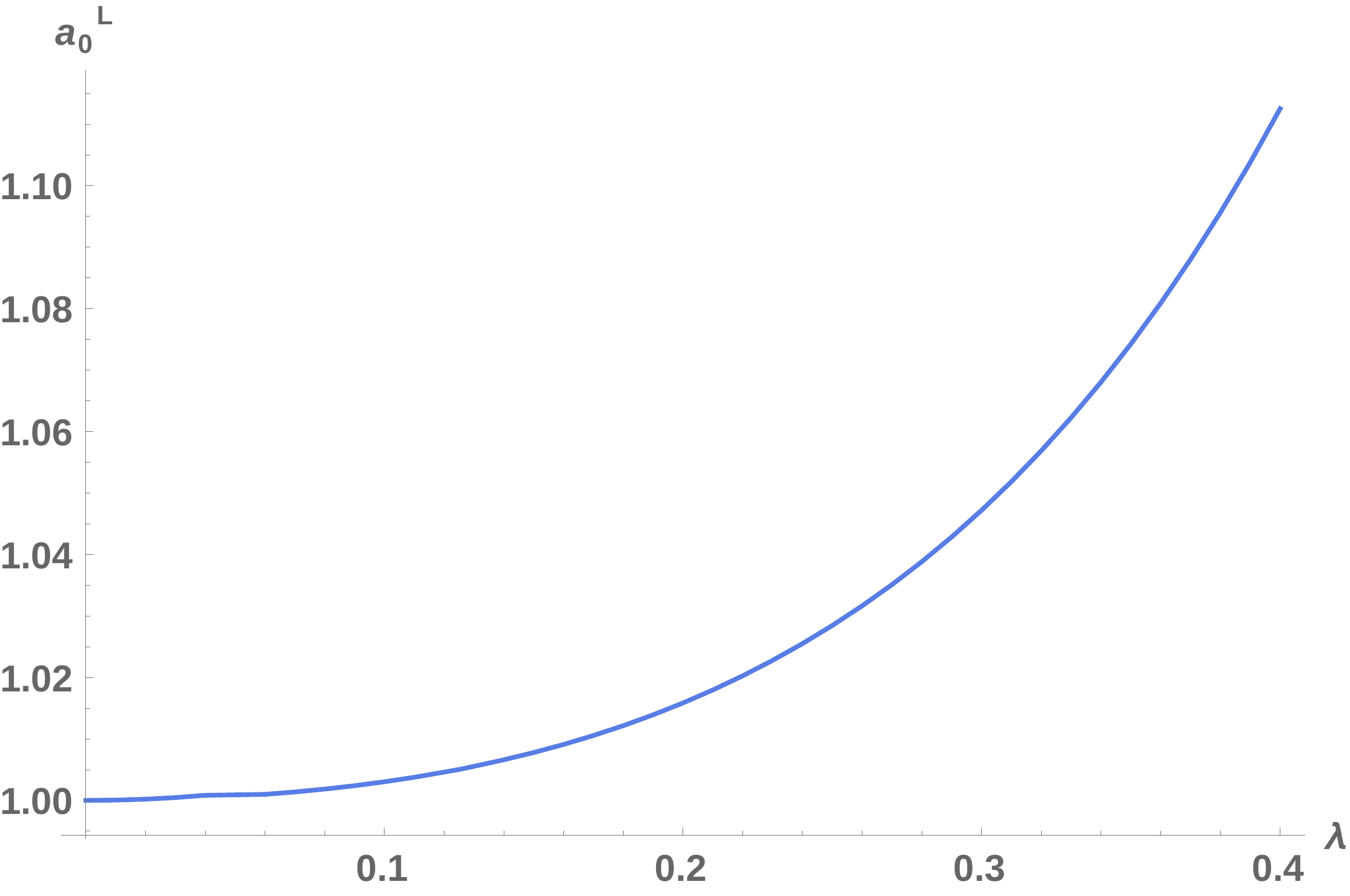}
        \caption{\centering{5 masses.}}
    \end{subfigure}
    \vspace{12mm}
    \begin{subfigure}[t]{0.49\textwidth}
        \includegraphics[width=\textwidth]{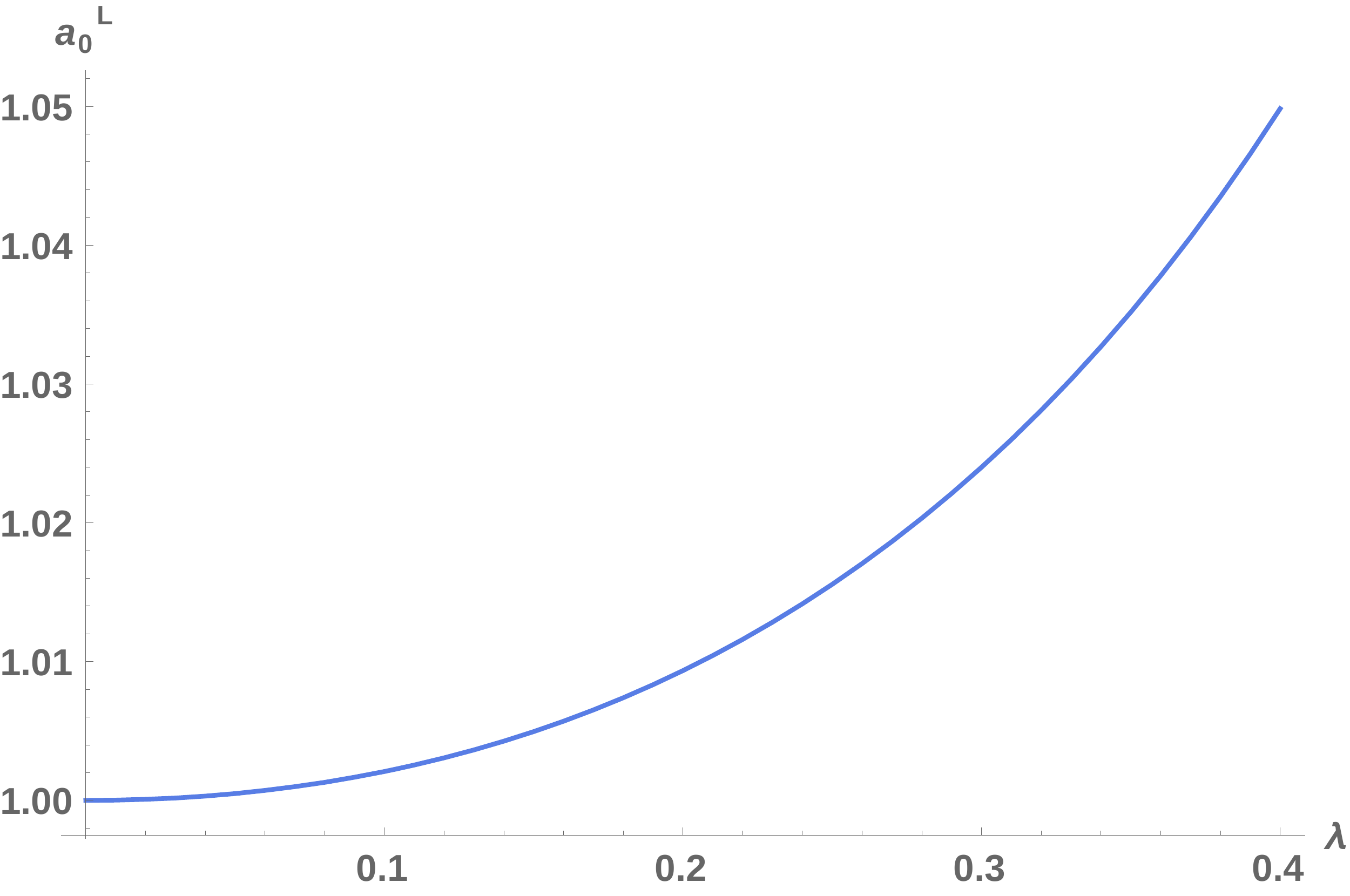}
           \caption{\centering{8 masses.}}
    \end{subfigure}
    \begin{subfigure}[t]{0.49\textwidth}
        \includegraphics[width=\textwidth]{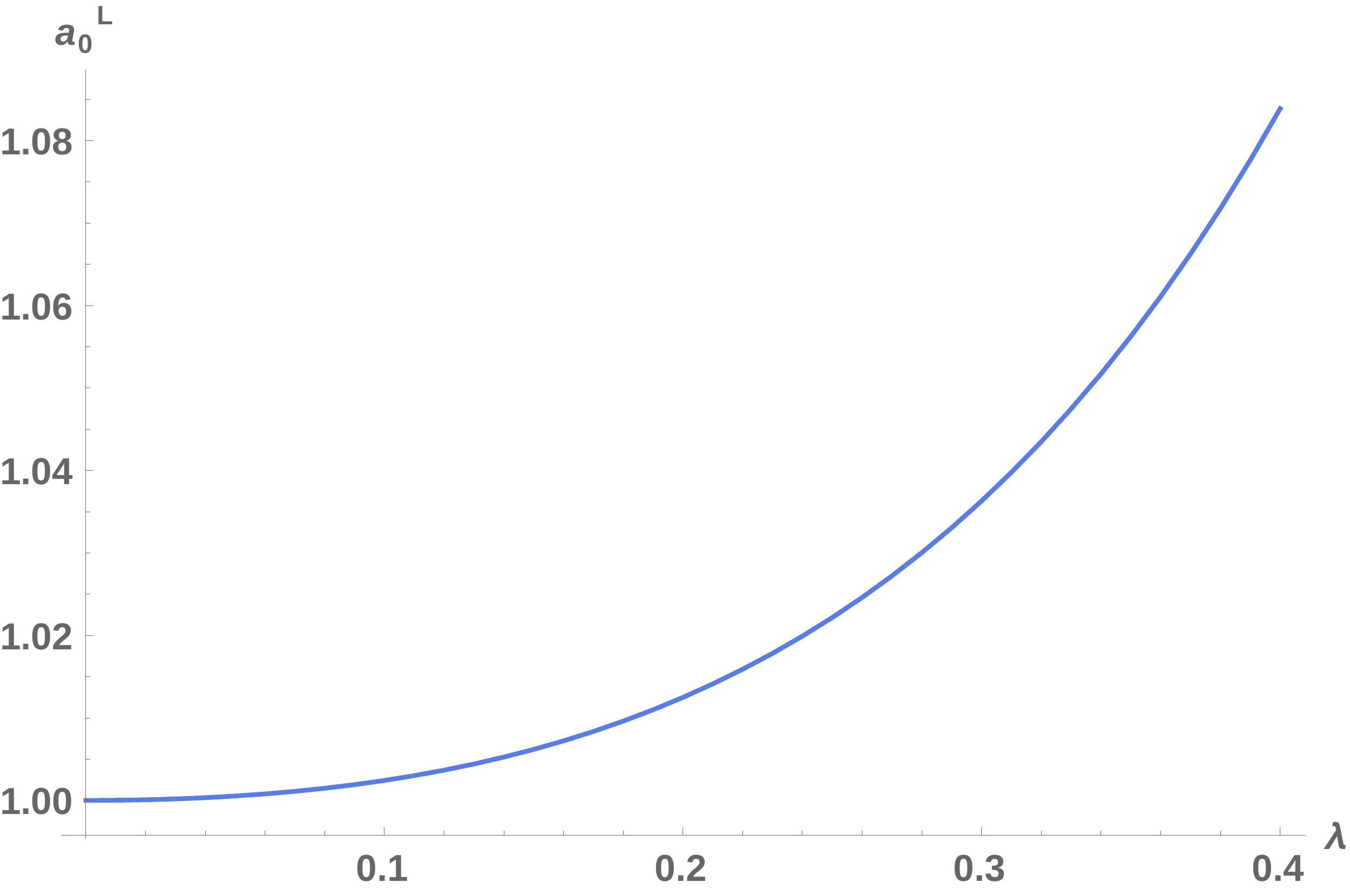}
           \caption{\centering{16 masses.}}
    \end{subfigure}
    \vspace{12mm}
    \begin{subfigure}[t]{0.49\textwidth}
        \includegraphics[width=\textwidth]{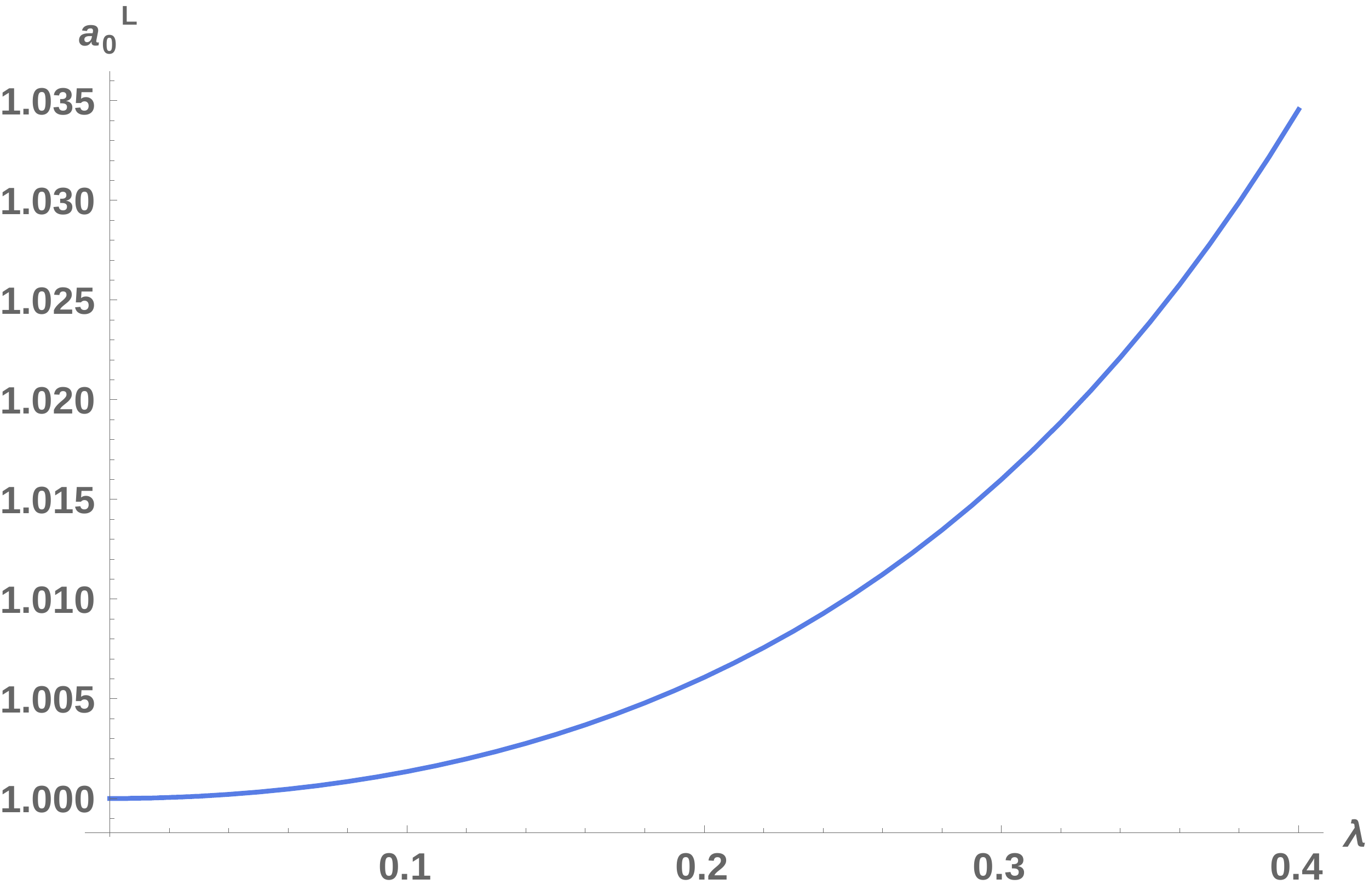}
           \caption{\centering{24 masses.}}
    \end{subfigure}
    \begin{subfigure}[t]{0.49\textwidth}
        \includegraphics[width=\textwidth]{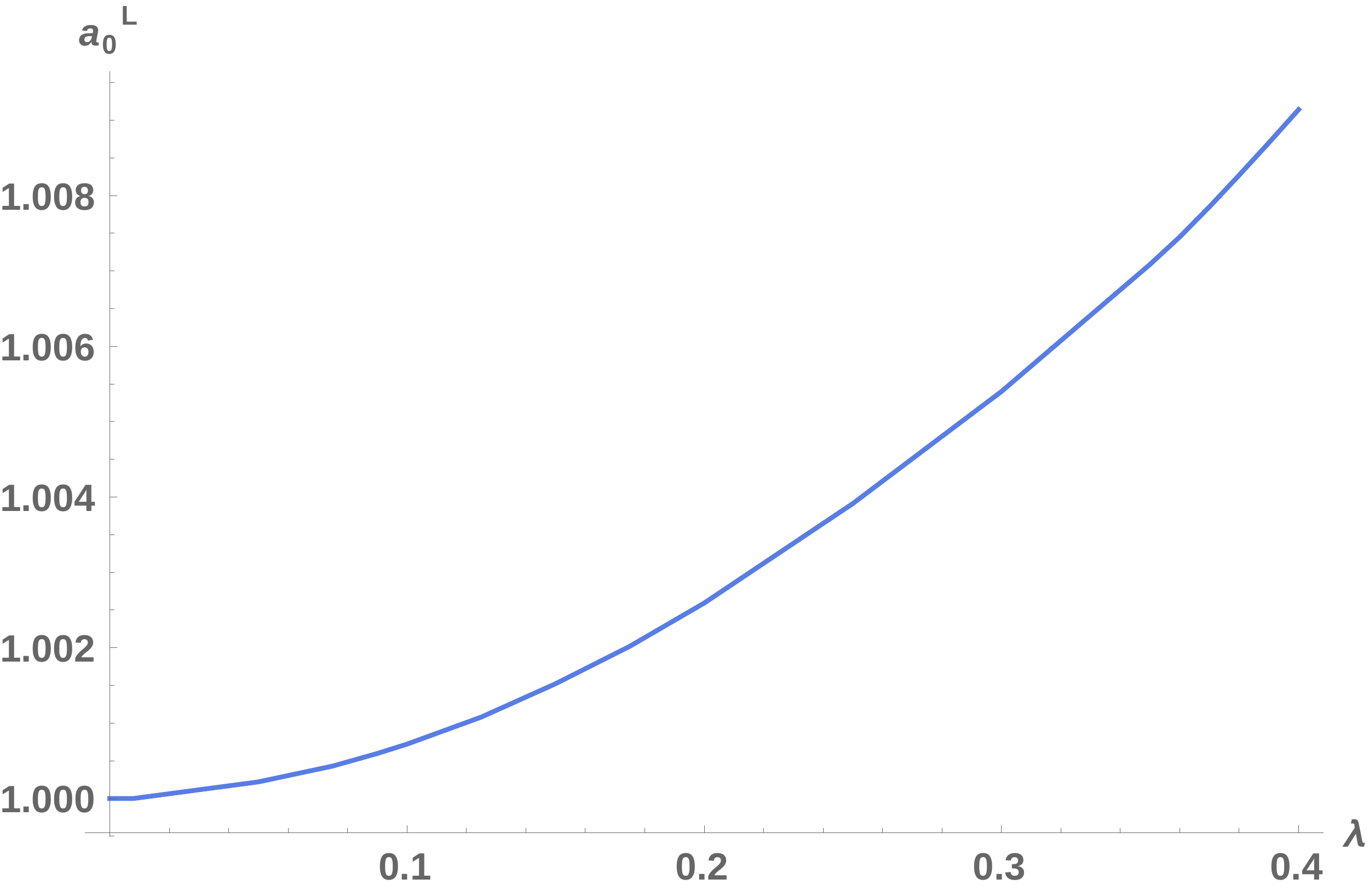}
           \caption{\centering{120 masses.}}
    \end{subfigure}
      \begin{subfigure}[t]{0.49\textwidth}
        \includegraphics[width=\textwidth]{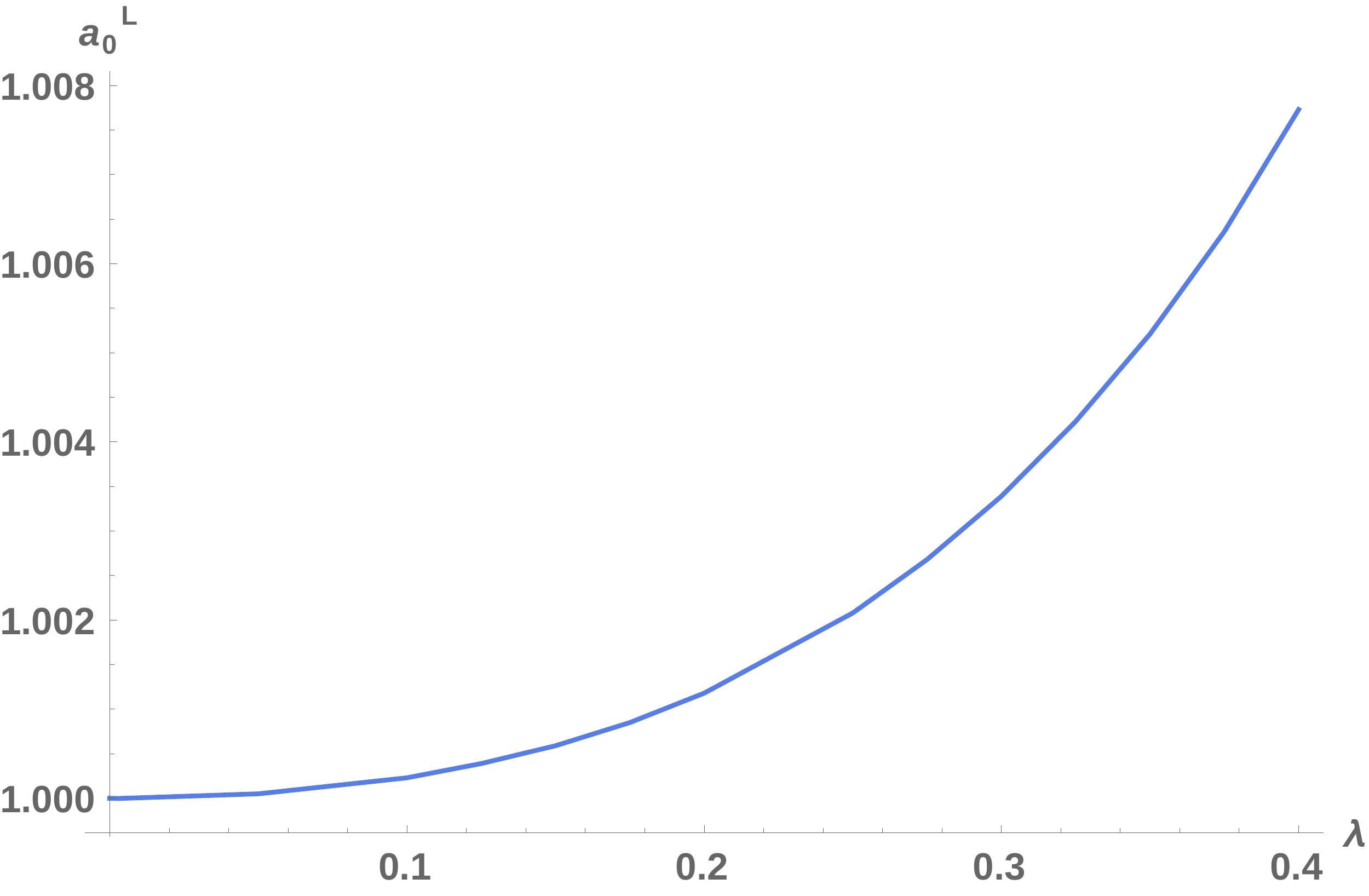}
           \caption{\centering{600 masses.}}
            \end{subfigure}
            \vspace{5mm}
     \caption{The scale factor of lattice cosmologies, $a_0^L$, as a function of $\lambda$ when the total energy is kept fixed (such that $\sum_i \tilde{m}_i$= constant). The values of the $\tilde{m}_i$ have been chosen so that $a_0^L=1$ when $\lambda=0$.}
    \label{fig:withint}
\end{figure}

In the previous section we compared our clustered lattice models to FLRW models with the same total proper mass, and the same total cluster mass. These results can, of course, be used to work out the difference in scale between lattice models with different values of $\lambda$, when they have the same $\sum_i m_i$ or $\sum_i \mathfrak{m}_i$. However, if we want to perform a similar analysis for lattice models with some fixed total bare mass $\sum_i \tilde{m}_i$ then a comparison with FLRW is less instructive. This is because there is a large difference between lattice models with some $\sum_i \tilde{m}_i$ and FLRW models that contain the same total bare mass in dust \cite{tim, matterincosmo}. What we can do, however, is compare between lattice models that have the same total bare mass but different values of $\lambda$. This allows us to see the consequences of structure formation on the large-scale cosmology if the total energy in the universe, including all proper masses and interaction energies, are kept fixed. The results are shown in Fig. \ref{fig:withint}.

We have chosen to normalise the scale factor by choosing bare mass parameters such that when $\lambda = 0$ we have $a^L_0=1$. Beyond this value, we find that the scale factor increases with $\lambda$. This growth is initially rather slow, but rapidly increases as $\lambda$ gets larger. Note that as the overall number of masses increases, the increase in the scale factor at large values of $\lambda$ becomes less pronounced. For example in the 5-cell lattice the scale factor is approximately $a^L_0 \simeq 1.05$ at $\lambda$ = 0.3, but for the 600-cell lattice it is only $\simeq 1.003$. This suggests that the effect of clustering decreases as the number of masses in the universe is increased, if we compare models with the same total energy, and drops below $\sim 1\%$ when the number of black holes is $\gtrsim 10^3$. This is qualitatively similar behaviour to the results shown in Fig. \ref{fig:scales2}, when the sum of intra-cluster interaction energies and proper masses were taken into account. It shows that the consequences of inter-cluster interactions are not particularly strongly affected by the clustering, which is in keeping with the near-constant nature of the green lines in Fig. \ref{fig:int}.

\section{Discussion}
\label{sec:conclude}

We have extended the black holes lattice cosmologies to include clusters of masses. The purpose of this has been to investigate the consequences of the existence of astrophysical structure on the large-scale properties of the Universe. The clusters themselves were created and parameterised in a controlled way, by splitting up the masses in the existing regular configurations and then gradually separating them. This gave us a set of four new black hole universes, containing $20$, $64$, $144$ and $2400$ masses, and with each model being controlled by a single parameter $\lambda$. We found that in addition to each black hole's own individual apparent horizons, that a collective horizon appeared around the clustered black holes when they were sufficiently close together (i.e. below a critical value of $\lambda$). This shared horizon disappears when $\lambda > \lambda_{\rm crit}$, and we have calculated explicit values for $\lambda_{\rm crit}$ for each of our four models.

By using and extending existing results from the literature, we derived expressions for the interaction energies between clusters, the proper mass of individual black holes, and the mass of clusters of black holes. We found that gravitational interaction energies within clusters diverges when the masses were brought together, but that the interaction energies between clusters stayed fairly constant over all values of $\lambda$. These results became increasingly significant when we started to compare our universes full of clustered black holes to FLRW models: we found that the most sensible way to compare these two classes of models was to include the intra-cluster interaction energies in the total energy budget, along with the proper masses of each of the black holes. Failure to include the interaction energies within clusters would appear to give deficits of order unity, if only the sum of proper masses of individual black holes were taken into account. The interaction energies within astrophysical structures therefore seem capable of having a non-negligible influence on the scale of the global cosmology.

In contrast, we found that when we also included interaction energies between clusters that the change in scale of the cosmology was relatively insensitive to the value of $\lambda$. This is despite the fact that the inter-cluster interaction energies were found to be very large in previous studies \cite{tim, matterincosmo}, and can be attributed to the result that large-scale interactions are not particularly affected by the degree of clustering in the model (see Fig. \ref{fig:int}). This leads to a somewhat puzzling conclusion; it appears we need to take interaction energies into account when calculating the total energy in our cosmological models, but if we take {\it all} interactions between {\it all} bodies into account then we get a total energy budget that is far exceeded by the total mass in a comparable dust-dominated FLRW model with the same scale factor.

The results presented here are a first step towards understanding the effects of structure formation on the global cosmology in exact solutions with clustered discrete masses. Further study will be required to determine how the different measures of energy and mass behave under dynamical evolution, which will almost certainly require the application of techniques from numerical relativity. If the total energy in the Universe is found to remain approximately constant, then we predict the effects of structure formation to be modest ($\sim 10\%$ when $N \sim 10$, decreasing to $\sim 1\%$ when $N\sim 10^3$). However, if only the proper mass of black holes remains constant then deviations of the order of $\sim 10\%$ seem possible, even when the number of masses is very large.

\paragraph{Acknowledgments:} JD and TC both acknowledge support from the STFC under grant STFC ST/N504257/1. We are grateful to E. Bentivegna, K. Rosquist and R. Tavakol for stimulating discussions about this project and the concepts of mass and energy in cosmology.

\end{document}